\documentclass{agujournal2019}

\usepackage[utf8]{inputenc}
\usepackage[T1]{fontenc}
\usepackage{url}
\usepackage{lineno}
\usepackage[inline]{trackchanges}
\usepackage{soul}
\usepackage{graphicx}
\usepackage{booktabs}
\usepackage{array}
\usepackage{longtable}
\usepackage{pdflscape}
\usepackage{multirow} 
\usepackage[table]{xcolor}
\usepackage{booktabs}
\usepackage{colortbl}
\usepackage{etoolbox}

\usepackage{xcolor}
\usepackage[colorlinks=true,
            citecolor=blue,
            linkcolor=blue,
            urlcolor=blue]{hyperref}
\definecolor{NavyHdr}{HTML}{1E3A5F}    
\definecolor{ExpHdr}{HTML}{EEF2FF}     
\definecolor{FocusBG}{HTML}{DBEAFE}    
\definecolor{AccentBlue}{HTML}{1D4ED8} 
\definecolor{DimText}{HTML}{A0AEC0}    
\definecolor{RuleLight}{HTML}{E8ECF1}  
\definecolor{RuleMid}{HTML}{334155}    
\definecolor{ExpBorder}{HTML}{C7D2FE}  
 
\newcommand{\dimcell}[1]{\textcolor{DimText}{#1}}
\newcommand{\focusbullet}{\textcolor{AccentBlue}{$\star$}}
\setkeys{Gin}{draft=false}

\draftfalse
\journalname{Journal of Advances in Modeling Earth Systems (JAMES)}

\begin{document}
\title{CORDEX-ML-Bench: A Benchmark for Data-Driven Regional Climate Downscaling---Experiment Design and Overview}
\authors{Neelesh Rampal\affil{1}, Jose González-Abad\affil{2}, Henry Addison\affil{3}, Jorge Baño-Medina\affil{2}, Maria Laura Bettolli\affil{4}, Valentina Blasone\affil{5}, Ben Booth\affil{6}, Erika Coppola\affil{5}, Serafina Di Gioia\affil{5}, Joshua Oldham-Dorrington\affil{7}, Antoine Doury\affil{8}, Francois Engelbrecht\affil{9}, Ramón Fuentes-Franco\affil{10}, Peter B. Gibson\affil{1}, Luca Glawion\affil{11}, Caroline Hardy\affil{9}, Mikhail Ivanov\affil{10}, Hugo Kyo Lee\affil{12}, Mikel N. Legasa\affil{13}, Matias Olmo\affil{14}, Andrew Orr\affil{15}, Julius Polz\affil{16}, Martin S. J. Rogers\affil{15}, Maybritt Schillinger\affil{17}, Shivani Sharma\affil{15}, Pedro M. M. Soares\affil{18}, Stefan Sobolowski\affil{7}, Jessica Steinkopf\affil{9}, Wenchang Tang\affil{5}, Jr-Ben Tian\affil{19}, Ricardo Tomé\affil{18}, Ko-Chih Wang\affil{19}, Yi-Chi Wang\affil{10}, Peter A. G. Watson\affil{3}, Tom Wetherell\affil{6}, Martin Widmann\affil{20}, José M. Gutiérrez\affil{2}}

\affiliation{1}{Earth Sciences New Zealand, Wellington, New Zealand}
\affiliation{2}{Instituto de Física de Cantabria (IFCA), CSIC-Universidad de Cantabria, Santander, Spain}
\affiliation{3}{School of Geographical Sciences, University of Bristol (UoB), Bristol, UK}
\affiliation{4}{Departamento de Ciencias de la Atmósfera y los Océanos, Universidad de Buenos Aires, CONICET, IFAECI/CNRS-IRD-UBA, Buenos Aires, Argentina}
\affiliation{5}{Abdus Salam International Centre for Theoretical Physics (ICTP), Trieste, Italy}
\affiliation{6}{Met Office, Exeter, UK}
\affiliation{7}{Geophysical Institute, University of Bergen and Bjerknes Centre for Climate Research (UiB), Bergen, Norway}
\affiliation{8}{Centre National de Recherches Météorologiques (CNRM), Université de Toulouse, Météo-France, CNRS, Toulouse, France}
\affiliation{9}{Global Change Institute, University of the Witwatersrand, Johannesburg, South Africa}
\affiliation{10}{Swedish Meteorological and Hydrological Institute, Rossby Centre (SMHI), Norrköping, Sweden}
\affiliation{11}{Institute of Meteorology and Climate Research -- Atmospheric Environmental Research (IMKIFU), Karlsruhe Institute of Technology, Campus Alpin, Garmisch-Partenkirchen, Germany}
\affiliation{12}{Jet Propulsion Laboratory (JPL), California Institute of Technology, Pasadena, CA, USA}
\affiliation{13}{Laboratoire des Sciences du Climat et de l'Environnement (LSCE-IPSL), CEA/CNRS/UVSQ, Université Paris-Saclay, Gif-sur-Yvette, France}
\affiliation{14}{Barcelona Supercomputing Center, Barcelona, Spain}
\affiliation{15}{British Antarctic Survey (BAS), Cambridge, UK}
\affiliation{16}{Institute of Meteorology and Climate Research -- Atmospheric Trace Gases and Remote Sensing (IMKASF), Karlsruhe Institute of Technology (KIT), Karlsruhe, Germany}
\affiliation{17}{Seminar for Statistics, ETH Zurich, Zurich, Switzerland}
\affiliation{18}{Instituto Dom Luiz (IDL), Faculdade de Ciências, Universidade de Lisboa, Lisbon, Portugal}
\affiliation{19}{Department of Computer Science and Information Engineering, National Taiwan Normal University (NTNU), Taipei, Taiwan}
\affiliation{20}{School of Geography, Earth and Environmental Sciences, University of Birmingham, Birmingham, UK}

\correspondingauthor{Neelesh Rampal}{neelesh.rampal@earthsciences.nz}
\begin{keypoints}
\item The first multi-domain benchmark for evaluating several leading data-driven downscaling algorithms.
\item Models trained only on historical periods underestimate future climate change signals, raising concerns for their broader use in downscaling.
\item Generative models generally outperform deterministic approaches for most metrics including extremes and capturing fine-scale spatial detail.
\end{keypoints}
\begin{abstract}

Machine learning (ML) has emerged as a cost-effective approach to complement dynamical downscaling for producing high-resolution regional climate projections. However, the absence of standardised training and evaluation protocols, applied consistently across multiple domains, continues to hinder meaningful model intercomparison. We introduce CORDEX-ML-Bench, a benchmark aligned with the Coordinated Regional Climate Downscaling Experiment (CORDEX), which constitutes the first phase of a community initiative to advance data-driven downscaling toward operational readiness, and complement future dynamical downscaling efforts under the Coupled Model Intercomparison Project Phase 7 (CMIP7). The framework targets downscaled daily maximum temperature and precipitation to $\sim$10 km resolution (20× increase) across three distinct pilot regions; European Alps, New Zealand, and Southern Africa. Using a perfect-model experimental design, we evaluate 40 ML configurations developed independently, spanning traditional ML, convolutional U-Nets, vision transformers, graph neural networks, and generative models based on diffusion, flow matching, and generative adversarial networks. Models are trained under two experimental periods, an empirical-statistical downscaling pseudo-reality (historical period only) and Emulator (historical and future periods) — and are evaluated against a core set of metrics developed specifically for assessing downscaling skill. Generative models consistently outperform deterministic approaches for precipitation, better capturing fine-scale variability and extremes. For maximum temperature, the generative advantage narrows and deterministic architectures remain competitive. Models trained solely on the historical period systematically underestimate future climate-change signals while those additionally trained on a future period perform considerably better. These findings raise concerns about historically trained models widely used in operational downscaling, underscoring the need for rigorous extrapolation testing.

\end{abstract}
\section*{Plain Language Summary}
High-resolution projections are traditionally obtained through dynamical downscaling, a process that enhances the spatial resolution of global climate models using computationally expensive physics-based regional climate models. Machine learning (ML) offers a cost-effective alternative for downscaling climate models, yet no consistent framework exists to evaluate these approaches, making it difficult to distinguish genuinely skilful methods from those that simply perform well on idealised tests. CORDEX-ML-Bench addresses this gap by providing a standardised benchmark that evaluates 40 ML downscaling configurations across three climatically diverse regions (European Alps, New Zealand, and Southern Africa), using a publicly available dataset, consistent experimental design, and common evaluation criteria applied to daily accumulated precipitation and maximum temperature. By providing open datasets and evaluation code, CORDEX-ML-Bench gives the community a shared infrastructure to rigorously develop the next generation of data-driven regional climate projections. Our results show that training on simulations spanning future periods, not just historical ones, is important for generating reliable climate projections, as models trained on historical data alone systematically underestimate future changes in temperature and precipitation extremes. Generative ML models, particularly diffusion and flow-matching approaches, generally outperform regression-based methods, especially for precipitation extremes. 

\section{Introduction}

Dynamical downscaling involves running a physics-based Regional Climate Model (RCM) over a limited domain at much higher spatial resolution than its driving Global Climate Model (GCM), constrained through lateral boundary conditions or nudging of selected large-scale variables. This allows them to explicitly represent mesoscale processes such as the interaction of atmospheric flow with complex orography, that are not resolved at the coarse spatial resolutions typical of GCMs ($\sim$130 km). RCMs are therefore important tools for studying climate variability and change at fine spatial scales, providing the detailed information needed for climate‑impact assessments and adaptation planning \citep[e.g.,][]{giorgi2015regional, rummukainen2016added}. An important international effort in this area is the Coordinated Regional Climate Downscaling Experiment \citep[CORDEX;][]{giorgi2009addressing}, which organizes multi-model dynamical downscaling across 14 domains covering most of the Earth's land areas. CORDEX provides substantially finer‑scale climate projections than GCMs—typically at 10–-25 km resolution in the Coupled Model Intercomparison Project Phase 6 \citep[CMIP6;][]{gutowski2016wcrp}. In RCMs, doubling the spatial resolution can increase the computational expense of simulations by roughly an order of magnitude, imposing heavy constraints on the achievable resolution in the context of domain size, the number of scenarios, and ensemble size \citep{kendon2025potential}. This subsequently limits the ability to comprehensively and robustly sample different types of uncertainty in regional climate projections. This is especially relevant for uncertainty as a result of internal variability — the irreducible uncertainty arising from chaotic processes within the climate system rather than external forcings. It represents a substantial and often underappreciated source of uncertainty in regional climate projections \citep{hawkins2009potential,maher_2021_large,deser2020insights,lehner_deser_2023,lewis2025generative}, particularly for extremes \citep{aalbers2018local,rampal2025downscaling}.

Machine learning (ML) is emerging as a complementary approach to dynamical downscaling, enabling high-resolution climate projections at a fraction of the cost and making it feasible to generate large ensembles \citep{rampal2024enhancing,sun2024deep,kendon2025potential}. More broadly, the computational efficiency of ML-based approaches may reduce barriers to generating high-resolution climate information in regions where dynamical downscaling at scale remains computationally prohibitive. Early efforts using convolutional neural networks \citep[e.g.,][]{bano_medina_suitability_2021,doury2023regional,van2023deep,soares2024high} produced the first ML approaches suitable for climate change downscaling, and have since contributed to a rapidly expanding body of literature, including several recent reviews \citep{rampal2024enhancing,sun2024deep,kendon2025potential}. Two main data-driven downscaling approaches have been extensively explored in the literature: empirical-statistical downscaling (ESD; also known as observational downscaling) and RCM emulation. Under the perfect prognosis approach, ESD develops empirical relationships between coarse-resolution atmospheric fields and local observations, with training constrained to the historical record \citep{maraun2010precipitation, gutierrez2019intercomparison}. In contrast, RCM emulation trains algorithms to replicate physics-based RCM output from coarse GCM-scale predictors, drawing on simulations that can span both historical and future climates \citep[e.g.,][]{doury2023regional, chadwick2011artificial, holden2015emulation, bano2024transferability, balmaceda2024use}. Once trained, either approach can be applied to GCM output to generate high-resolution projections \citep{rampal2024enhancing}.

Although climate downscaling and computer vision (image super-resolution) share enough parallels to motivate similar model architectures, there are several important differences. Most notably, climate downscaling requires predictions that are physically consistent, and must generalise reliably across weather systems and climate states well outside the training distribution \citep{rampal2024enhancing, maraun2015value}. An important limitation for ML downscaling is the lack of coordination and standardisation. Studies frequently differ in their choice of predictor variables, training strategy, experiment design, evaluation metrics, and geographic domains, making systematic intercomparison of leading ML models difficult \citep{gonzalez2025deep,harder2026benchmarking}. Many metrics borrowed from computer science are poorly suited to climate applications, and standard training and evaluation protocols in the downscaling literature are often not designed with real-world use cases in mind \citep{rampal2024enhancing}. Without clearly defined standards, it remains difficult to distinguish genuine methodological advances from performance gains attributable to favourable experimental design. Benchmarking frameworks have proven transformative in related fields. WeatherBench \citep{rasp2020weatherbench,rasp2024weatherbench} and AIMIP \citep{henn2026aimip} are notable examples, establishing standardised protocols for medium-range weather prediction and measurably accelerated progress in data-driven forecasting. The need for robust benchmarking frameworks specific to climate downscaling has been recognised \citep{langguth2024benchmark,harder2026benchmarking}, but existing efforts remain poorly aligned with the evaluation needs of the regional climate modelling community. Unlike weather forecasting benchmarks, such frameworks must address challenges specific to climate projections: extrapolation to future climates across GCMs, transferability across emissions scenarios, and skill across diverse geographic domains. Despite the growing body of downscaling literature, these considerations are addressed by only a small number of studies, underscoring the need for a comprehensive benchmarking framework tailored to the demands of climate projection applications.

To address these gaps, we introduce CORDEX-ML-Bench, a standardised benchmarking framework and dataset for data-driven climate downscaling explicitly aligned with CORDEX, openly available at \citet{Rampal2026_CORDEX_ML_Bench_companion}. CORDEX-ML-Bench supports rigorous and reproducible evaluation of two main data-driven downscaling approaches; ESD and RCM emulation, and builds on lessons learned from VALUE \citep{gutierrez2019intercomparison} -- a previous European collaboration to benchmark statistical techniques for downscaling, while incorporating recent advances in ML and explicitly targeting RCM emulation. The framework aims to foster coordinated methodological development, establish standards of practice, and encourage researchers to develop and evaluate new solutions, ultimately laying the foundation for integrating data-driven downscaling into operational climate-projection workflows. This first phase of the benchmark places particular emphasis on experiments and evaluation metrics that address key climate-specific challenges such as extrapolation to future climates, non-stationarity, and cross-model and cross-domain transferability. This initial benchmark targets the downscaling of two key surface variables (daily precipitation and maximum temperature) to approximately 10 km ($\sim$0.11$^\circ$) resolution at daily frequency, using predictor fields at 2$^\circ$ ($\sim$200 km)---an effective 20$\times$ spatial resolution increase.

The target resolution was chosen to be consistent with commonly used CORDEX outputs. These two variables were prioritized as they are widely used surface diagnostics in climate-impact assessments, and because they present contrasting predictability characteristics. This effort has been facilitated by the CORDEX Machine Learning Task Force, which has subsequently evolved to become a ML Task Team \citep{gutierrez_etal_inpress}. The benchmark encompasses standardised datasets for three climatically and geographically diverse pilot regions — the European Alps, New Zealand, and South Africa. In this first phase, all algorithms are trained in a perfect model framework, in which coarse-resolution predictors are derived by spatially coarsening RCM atmospheric fields to a typical GCM resolution, with the original high-resolution RCM fields serving as targets. The imperfect framework, in which models are trained using GCM fields directly as predictors \citep[analogous to boundary conditions or spectral nudging in dynamical downscaling;][]{van2023deep, rampal2024enhancing, doury2023regional}, introduces additional complexity due to systematic GCM–RCM state discrepancies, which make the statistical relationships harder to learn \citep{bano2024transferability}. The perfect-model framework is therefore adopted here to isolate the downscaling function from these additional sources of error \citep{bano2024transferability}.

\section{Materials and Methods}

Unlike weather forecasting benchmarks such as WeatherBench \citep{rasp2020weatherbench}, where day-to-day accuracy is of primary concern, climate downscaling evaluation must additionally account for distributional skill, long-term trends, non-stationarity, and extrapolation to future climate conditions. We acknowledge that no single metric or set of metrics can fully characterise model quality for all use cases. Rather than treating CORDEX-ML-Bench as a challenge with a single leaderboard, we therefore present it as a framework for systematic model comparison, one that establishes a core set of evaluation criteria that should be examined before any ML-based approach can be trusted to produce climate projections.

We evaluate model performance across four generalisation settings: (i) cross-validation over the historical period (in-sample performance); (ii) interpolation to an unseen period (RCM emulation); (iii) extrapolation to future out-of-distribution conditions (ESD); and (iv) transferability to unseen real-world CMIP GCMs within the perfect-model setting. Aspect (iv) was incorporated into the experimental design but will be the focus of a subsequent study. Evaluation metrics were developed in consultation with both climate scientists and ML researchers to reflect the unique requirements of climate applications. The core evaluation metrics encompass historical climatological skill, the ability to reproduce climate change signals, and performance across the full distribution of simulated values, with particular attention to extremes.

\subsection{Overview of Benchmarking Dataset}\label{sec:overview}

Unlike data‑driven weather forecasting—where the goal is to predict the next atmospheric state from the previous one—climate downscaling predicts the high‑resolution field ($y$) at time $t$ directly from coarse‑resolution predictors ($X$) at the same time \textit{t}. 

\begin{figure}
\includegraphics[width=1.1\textwidth]{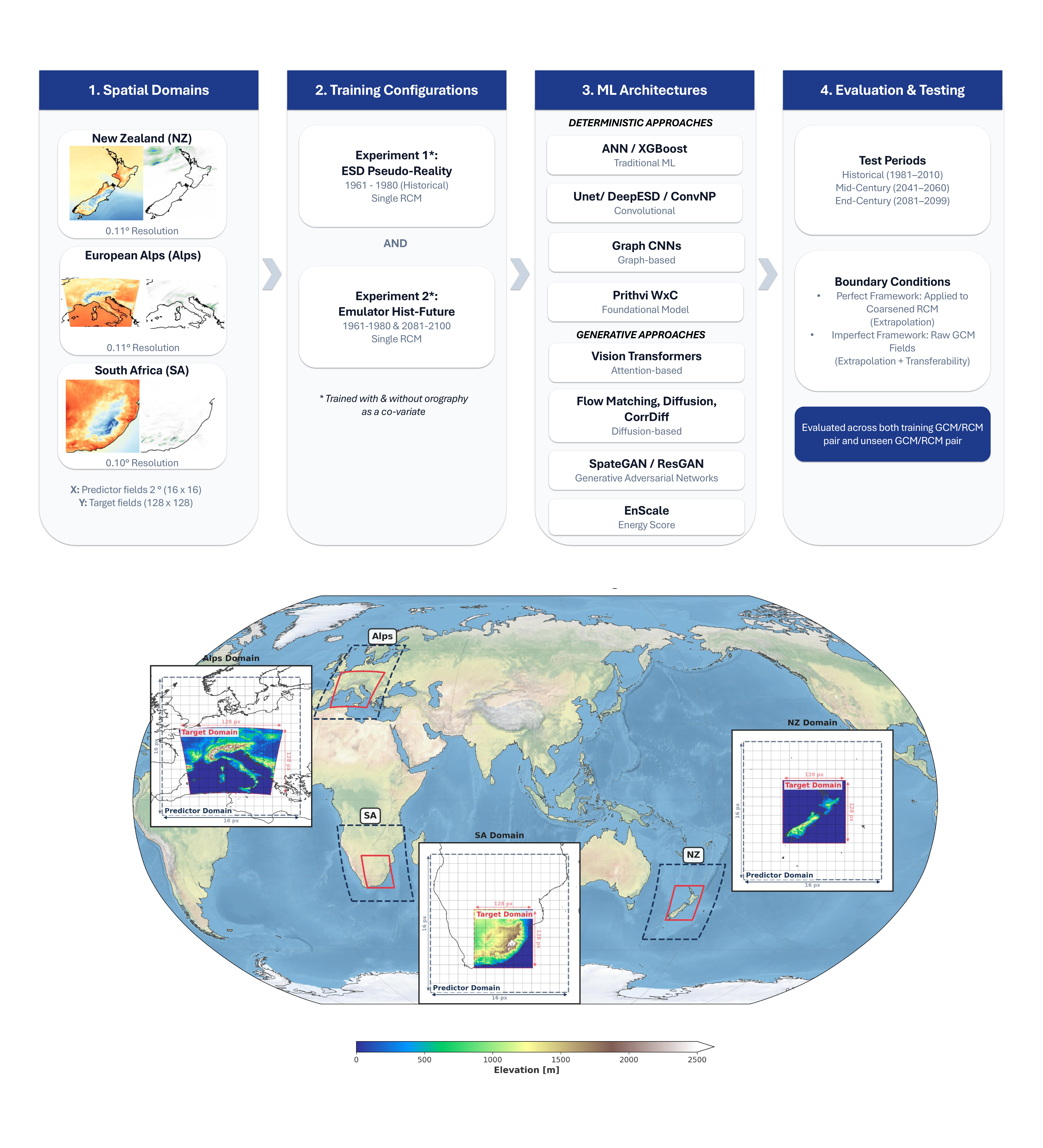}
\caption{CORDEX-ML-Bench Experimental Design. The top panel provides a schematic overview of the benchmark, summarizing the spatial domains, training period/experiment, machine-learning architectures, and evaluation periods. The bottom panel shows the regional domains in greater detail, including the coarse-resolution predictor domains (dashed black lines), the nested high-resolution target domains (solid red lines), and the underlying surface elevation (color scale in meters). Insets highlight each region with predictor and target grids overlaid on regional orography.}
\label{fig:design}
\end{figure}

Training and evaluation datasets span three geographically distinct regions: the European Alps and New Zealand represent mid‑latitude regions strongly influenced by complex orography, while South Africa provides a more subtropical domain shaped by tropical and mid‑latitude weather systems. The benchmark targets two high-resolution variables simulated by the RCM: daily maximum temperature (\textit{tasmax}) and daily accumulated precipitation (\textit{pr}). Target fields are provided on a 128 $\times$ 128 grid at approximately 10 km resolution (0.11$^\circ$ for the Alps and New Zealand domains; 0.10$^\circ$ for South Africa), as illustrated for the three regional domains in Figure~\ref{fig:design}. Predictor fields are coarsened directly from the high-resolution RCM output using the perfect framework, ensuring spatial and temporal alignment between the low- and high-resolution fields used for training. The daily predictor variables used in this study are listed in Table~\ref{tab:cordex_variables}. They were selected based on their availability in CMIP6 output, their widespread use in the perfect-prognosis downscaling literature \citep{maraun2015value,gutierrez2019intercomparison}, and their direct applicability to CMIP GCM projections. Predictor fields are coarsened to a 2$^\circ$ resolution using conservative remapping, yielding a 16 $\times$ 16 grid across all three regions. In a few cases, the 850 hPa predictors contain missing values over high-elevation areas. These affect a small fraction of timesteps, which are masked out when computing the evaluation metrics. The predictor domain intentionally covers a larger region than the target domain to prevent information scarcity about the domain edges. Grid dimensions of 16 $\times$ 16 and 128 $\times$ 128 (both powers of two) were chosen to ensure compatibility with standard deep learning architectures, while the factor-of-8 spatial ratio between predictor and target grids facilitates the direct application of common upsampling approaches. 

\definecolor{lightgray}{gray}{0.95}
\begin{table}[h]
\centering
\renewcommand{\arraystretch}{1.4}
\caption{Overview of predictor and target variables in the CORDEX-ML-Bench dataset.}
\label{tab:cordex_variables}
\begin{tabular}{l l p{3.5cm} l l}
\toprule
\rowcolor{lightgray}
\textbf{Long name} & \textbf{Short name} & \textbf{Description} & \textbf{Unit} & \textbf{Levels} \\
\midrule
\rowcolor{lightgray}
\multicolumn{5}{l}{\textbf{Predictor fields (2°)}} \\
Geopotential        & \textit{z} & Height of a pressure level     & m$^2$ s$^{-2}$ & 850, 700, 500 \\ \hline
Temperature         & \textit{t} & Atmospheric temperature        & K              & 850, 700, 500 \\ \hline
Specific humidity   & \textit{q} & Water vapour mixing ratio      & kg kg$^{-1}$  & 850, 700, 500 \\ \hline
U-component of wind & \textit{u} & Zonal wind speed               & m s$^{-1}$    & 850, 700, 500 \\ \hline
V-component of wind & \textit{v} & Meridional wind speed          & m s$^{-1}$    & 850, 700, 500 \\
\midrule
\rowcolor{lightgray}
\multicolumn{5}{l}{\textbf{Target fields ($\sim$10\,km)}} \\
Maximum temperature & tasmax & Daily maximum near-surface temperature & K            & 2-metre \\ \hline
Precipitation       & pr     & Daily accumulated precipitation        & mm day$^{-1}$ & Surface \\
\midrule
\rowcolor{lightgray}
\multicolumn{5}{l}{\textbf{Constants}} \\
Orography           & orog   & Surface elevation                      & m             & Surface \\
\bottomrule
\end{tabular}
\end{table}

For each region, training employs a single RCM simulation forced by a single GCM under two different periods: (i) a model trained exclusively on the historical period (1961–1980) within the ESD pseudo‑reality setting, and (ii) the RCM Emulator setting where a model is trained on the combined historical (1961-1980) and future periods (2080-2099). Further details of these experiments are provided in Section \ref{sec:exp_design}. Evaluation is performed using both the driving GCM—over a period not included in training—and an entirely independent “unseen” GCM. Models are evaluated in two settings: the perfect framework, consistent with the training setup, and the imperfect setting, in which models are applied directly to GCM fields (reflecting real-world application to CMIP-style outputs). Note that "imperfect evaluation" refers solely to the evaluation setting, not to the imperfect training framework discussed earlier. Both frameworks are discussed in detail in \citep{kendon2025potential, rampal2024enhancing, boe2023simple}. For this benchmark, all training and evaluation data are provided at daily temporal resolution to reduce dataset size and ensure accessibility across a wide range of computing environments. This choice also allows us to begin with a simpler setting before extending the methodology to more challenging temporal resolutions (e.g., sub-hourly). The full dataset—covering all training and testing splits across the three regions—amounts to approximately 30GB and is supplied as regional NetCDF files in compressed archives. The data are available for download on Zenodo \citep{Rampal2026_CORDEX_ML_Bench_companion}.

\subsection{Experimental Protocol}\label{sec:exp_design}

\subsubsection{RCM Training Simulations}

As introduced in Section \ref{sec:overview}, the training data used in this study are derived from previously produced dynamical downscaling regional efforts for CMIP5 and CMIP6. Different RCM configurations are used for each region in this benchmark. For the European Alps, target fields are generated using the ALADIN63 RCM \citep{nabat2020modulation}, a conventional limited-area RCM driven by the CNRM-CM5 GCM \citep{voldoire2013cnrm} under the CMIP5 framework. For South Africa and New Zealand, target fields are produced using the Conformal Cubic Atmospheric Model \citep[CCAM;][]{mcgregor2008updated,thatcher2009using}, a global non-hydrostatic variable-resolution climate model. CCAM employs spectral nudging to constrain the large-scale circulation toward that of the driving GCMs \citep{chapman2023evaluation,gibson2023high, engelbrecht2025extreme}. For both the Southern Africa and New Zealand domains, the primary training RCM (CCAM) is driven by the ACCESS-CM2 GCM from CMIP6. Unlike limited-area RCMs such as ALADIN, CCAM operates on a global mesh with locally refined resolution over the target domains, offering a distinct and well-evaluated approach to regional downscaling \citep{gibson2024dynamical,gibson2025downscaled,truong2025simulation,engelbrecht2025extreme, campbell2024comparison, goddard2025high}.

Each of these RCM configurations has been independently evaluated against observations and reanalysis products, demonstrating skill in reproducing regional climatological means, interannual variability, and climate change responses across the respective target domains \citep{nabat2020modulation,campbell2024comparison,engelbrecht2025extreme}. These simulations therefore serve as physically consistent and well-characterised references for key aspects of regional climate, including historical temperature and precipitation climatology, seasonal and interannual variability, and projected climate change signals. We acknowledge that all RCM simulations carry systematic biases inherited from both the driving GCM and the RCM configuration. It is important to emphasise that the goal of this benchmark is not to assess performance against observations directly, but rather to assess the ability of ML models to emulate the RCM --- including its biases --- thereby treating the dynamical downscaling output as the learning target. 

\subsubsection{Experimental Design and Evaluation}

\begin{figure}
\noindent\includegraphics[width=1.2\textwidth]{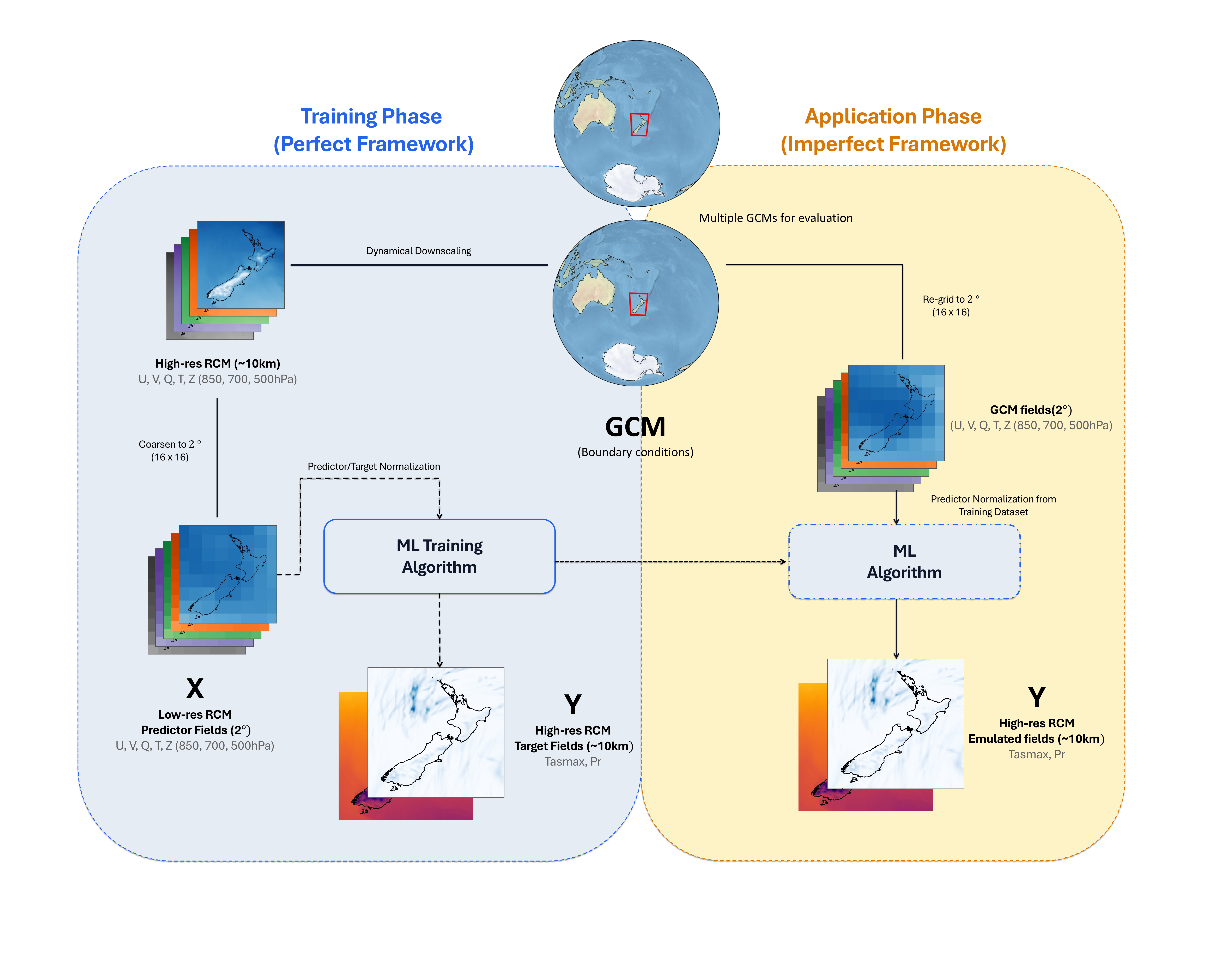}
\caption{Schematic of the perfect (training phase, left) and imperfect (application phase, right) frameworks used in CORDEX-ML-Bench. In the perfect framework, high-resolution RCM fields are coarsened to produce low-resolution predictor fields (X), which are paired with high-resolution RCM target fields (Y) for training. RCM-derived target fields (Y) are used for evaluation in both the perfect and imperfect settings — the key distinction being that in the imperfect evaluation setting (right), the trained model receives GCM fields as predictors rather than RCM-derived fields, representing real-world CMIP-style application. Note that "imperfect" here refers to evaluation only, not to imperfect training.}

\label{fig:perfect-imperfect}
\end{figure}

In the ESD experiment, models are trained on the historical period (1961--1980) using the RCM simulation as a pseudo-reality of observations, allowing out-of-distribution performance to be evaluated against a known future — something not possible when testing against true observational data \citep{vrac2007general}. Future projections in CORDEX are driven by emission scenarios, expressed as Representative Concentration Pathways (RCPs) or Shared Socioeconomic Pathways (SSPs) depending on the driving GCM. In the RCM emulator experiment, the training dataset is extended to 40 years by adding end-of-century simulations under a high-emission scenario (2080–2099; RCP8.5 for the European Alps, SSP3-7.0 for South Africa and New Zealand), exposing models to a broader range of climate states and shifting the evaluation emphasis from extrapolation to interpolation. Both experiments are conducted with and without high‑resolution surface elevation to systematically quantify the influence of static geographic predictors. All models are trained within the perfect framework (Figure~\ref{fig:perfect-imperfect}, left), as this avoids the GCM–RCM discrepancies that degrade out-of-sample transferability when models are trained on raw GCM fields \citep{doury2024suitability, bano2024transferability}.

Once trained, models are applied to coarsened RCM fields (perfect framework) and raw GCM predictors (imperfect evaluation), the latter representing a typical operational setting (Figure~\ref{fig:perfect-imperfect}, right). This paper focuses exclusively on the perfect framework, evaluating performance on two unseen periods from the training GCM: a historical cross-validation period (1981–2010) and mid-century conditions (2041–2060; Table~\ref{tab:cordex-ml-bench}). This perfect-model framework is well established in the ML downscaling and emulation literature \citep[e.g.,][]{bano2024transferability,doury2023regional,rampal2024enhancing,addison2026machine, balmaceda2024use, chadwick2011artificial, maraun2015value}, and provides a controlled setting in which model skill can be attributed unambiguously to architectural and training choices rather than to observational uncertainty or RCM error. 


\definecolor{headerblue}{RGB}{220,220,220}
\definecolor{lightline}{RGB}{200,200,200}

\begin{table}[ht!]
\centering
\setlength{\tabcolsep}{5pt}
\renewcommand{\arraystretch}{1.3}
\small
\caption{Summary of the CORDEX-ML-Bench experimental design and evaluation
protocol. \textbf{A)} Regional modelling chains and independent out-of-sample
GCMs for transferability testing. \textbf{B)} Structured evaluation matrix
isolating pure downscaling skill from structural biases.
\colorbox{FocusBG}{$\star$~Highlighted rows} indicate the configurations
evaluated in this paper (training-GCM boundary conditions; periods 1981--2010
and 2041--2060). Remaining rows (greyed) are defined for completeness and future work.}
\label{tab:cordex-ml-bench}

\begin{tabular}{p{2.8cm} p{1.8cm} p{4.0cm} p{4.0cm}}
\toprule
\rowcolor{NavyHdr}
\multicolumn{4}{l}{\textcolor{white}{\textbf{A\enspace Regional Domains and RCMs}}} \\
\rowcolor{white}
\textbf{Region} & \textbf{RCM} &
\textbf{Training GCM} & \textbf{Test GCM (out-of-sample)} \\
\arrayrulecolor{RuleMid}\midrule\arrayrulecolor{black}
\textbf{European Alps} & ALADIN63 &
  CNRM-CM5 \textit{\small(CMIP5)} & MPI-ESM-LR \textit{\small(CMIP5)} \\
\arrayrulecolor{RuleLight}\hline\arrayrulecolor{black}
\textbf{South Africa}  & CCAM   &
  ACCESS-CM2 \textit{\small(CMIP6)} & NorESM2-MM \textit{\small(CMIP6)} \\
\arrayrulecolor{RuleLight}\hline\arrayrulecolor{black}
\textbf{New Zealand}   & CCAM   &
  ACCESS-CM2 \textit{\small(CMIP6)} & EC-Earth3 \textit{\small(CMIP6)} \\
\bottomrule
\end{tabular}
 
\vspace{1.2em}
 
\begin{tabular}{p{0.6cm} p{3.8cm} p{3.0cm} p{3.6cm} p{3.0cm}}
\toprule
\rowcolor{NavyHdr}
\multicolumn{5}{l}{\textcolor{white}{\textbf{B\enspace Training and Evaluation Matrix}}} \\
\rowcolor{white}
  & \textbf{Test configuration}
  & \textbf{Evaluation period}
  & \textbf{Predictor source}
  & \textbf{Driving GCM} \\
\arrayrulecolor{RuleMid}\midrule\arrayrulecolor{black}
 
\rowcolor{ExpHdr}
\multicolumn{5}{l}{%
  \textbf{Experiment 1 \textemdash{} ESD}\quad
  \footnotesize Train: 1961--1980 (Historical)} \\
\arrayrulecolor{ExpBorder}\hline\arrayrulecolor{black}
 
\rowcolor{FocusBG}
\focusbullet & Perfect cross-validation
  & \textbf{1981--2010}
  & RCM (coarsened) & Training \\
\arrayrulecolor{RuleLight}\hline\arrayrulecolor{black}
 
\dimcell{2} & \dimcell{Imperfect cross-validation}
  & \dimcell{1981--2010}
  & \dimcell{GCM (raw)} & \dimcell{Training} \\
\arrayrulecolor{RuleLight}\hline\arrayrulecolor{black}
 
\rowcolor{FocusBG}
\focusbullet & Perfect extrapolation
  & \textbf{2041--2060}
  & RCM (coarsened) & Training \\
\arrayrulecolor{RuleLight}\hline\arrayrulecolor{black}
 
\dimcell{4} & \dimcell{Imperfect extrapolation}
  & \dimcell{2041--2060; 2080--2099}
  & \dimcell{GCM (raw)} & \dimcell{Training} \\
\arrayrulecolor{RuleLight}\hline\arrayrulecolor{black}
 
\dimcell{5} & \dimcell{Perfect extrapolation (transfer)}
  & \dimcell{2041--2060; 2080--2099}
  & \dimcell{RCM (coarsened)} & \dimcell{Out-of-sample} \\
 
\arrayrulecolor{RuleMid}\midrule\arrayrulecolor{black}
 
\rowcolor{ExpHdr}
\multicolumn{5}{l}{%
  \textbf{Experiment 2 \textemdash{} RCM Emulator}\quad
  \footnotesize Train: 1961--1980 \& 2080--2099 (Emulator Hist--Future)} \\
\arrayrulecolor{ExpBorder}\hline\arrayrulecolor{black}
 
\rowcolor{FocusBG}
\focusbullet & Perfect cross-validation
  & \textbf{1981--2010}
  & RCM (coarsened) & Training \\
\arrayrulecolor{RuleLight}\hline\arrayrulecolor{black}
 
\dimcell{2} & \dimcell{Imperfect cross-validation}
  & \dimcell{1981--2010}
  & \dimcell{GCM (raw)} & \dimcell{Training} \\
\arrayrulecolor{RuleLight}\hline\arrayrulecolor{black}
 
\rowcolor{FocusBG}
\focusbullet & Perfect interpolation
  & \textbf{2041--2060}
  & RCM (coarsened) & Training \\
\arrayrulecolor{RuleLight}\hline\arrayrulecolor{black}
 
\dimcell{4} & \dimcell{Imperfect interpolation}
  & \dimcell{2041--2060}
  & \dimcell{GCM (raw)} & \dimcell{Training} \\
\arrayrulecolor{RuleLight}\hline\arrayrulecolor{black}
 
\dimcell{5} & \dimcell{Perfect interpolation (transfer)}
  & \dimcell{2041--2060; 2080--2099}
  & \dimcell{RCM (coarsened)} & \dimcell{Out-of-sample} \\
\arrayrulecolor{RuleLight}\hline\arrayrulecolor{black}
 
\dimcell{6} & \dimcell{Imperfect interpolation (transfer)}
  & \dimcell{2041--2060; 2080--2099}
  & \dimcell{GCM (raw)} & \dimcell{Out-of-sample} \\
 
\bottomrule
\end{tabular}
 
\medskip
 
\end{table}

\subsection{Contributing Models}


CORDEX-ML-Bench encompasses 21 distinct ML model architectures (spanning 40 different configurations) developed independently by 13 research institutions, covering approaches from standard neural networks and gradient-boosted trees to state-of-the-art generative models based on diffusion, flow matching, and Generative Adversarial Networks (GANs).  This selected subset is broadly representative of the current literature and reflects the state-of-the-art in ML-based statistical downscaling. This diversity is intentional, enabling the community to assess the relative merits of a wide range of methods within a consistent evaluation framework applied to multiple regions. To reflect a more realistic training and evaluation setting, participating groups were provided only with predictor fields for the test period, with target values withheld until independent evaluation. This ensured a more robust and independent evaluation by preventing participating groups from overly tuning their models to the test set. Training strategies varied across contributing models, including validation splits and stopping criteria --- some models trained for a fixed number of epochs while others used validation loss or other metrics for checkpoint selection --- all of which may influence the interpretation of relative performance.

Contributing models are divided into two categories: deterministic models, which produce a single prediction per time step, and generative models, which produce stochastic ensembles sampling from the conditional distribution of the predictand. For most experiments, groups developed paired configurations---one with high-resolution orography as a static co-variate and one without---to systematically assess the value of topographic information. Configurations including orography carry the suffix \textit{-orog}. The majority of models were trained independently for each target variable, though a small number of models were designed to predict both variables jointly. Further details, including hyperparameters and hardware specifications, are provided in Supplementary Tables T1-T3. Reported training times are approximate, per configuration, and correspond to the RCM Emulator experiment trained over 40 years of simulation.

\subsubsection{Deterministic Models}

The deterministic algorithms range from shallow networks to sophisticated deep learning architectures. Most commonly, models used Mean Squared Error (MSE) in the loss functions for temperature and implemented more tailored loss functions for precipitation, reflecting the zero-inflated and heavy-tailed nature of daily rainfall distributions. An overview of each algorithm is provided, with model names followed by the contributing institution or consortium in square brackets:
\begin{itemize}

\item \textbf{Rossby-UNet [SMHI]} is adapted from \citet{fuentes2025pan} and is a five-layer U-Net \citep{ronneberger2015u}, with a Convolutional Block Attention Module \citep[CBAM;][]{woo2018cbam}, Group normalisation, and Sigmoid Linear Unit (SiLU) activations, trained with a combined MSE-Fast Fourier Transform (FFT) loss that penalises errors in both spatial and spectral domains equally. Predictors are standardised using local monthly climatologies and bicubically interpolated to the target grid. 

\item \textbf{DetUNet and DetUNet-v2 [Earth Sciences NZ]} are adapted from \citet{rampal2025reliable, ward2025intercomparison}: a $\sim$3M-parameter U-Net with residual convolutions, self-attention at the latent stage, and Feature-wise Linear Modulation (FiLM) layers for day-of-year conditioning, trained with an MSE loss for 250 epochs. Precipitation is log-normalised prior to training; temperature is z-score standardised at the grid-point level. The two variants differ in learning rate (smaller in -v2) and activation function: DetUNet-v2 uses LeakyReLU(0.5) for temperature.

\item \textbf{CNRM-UNeT [CNRM]} has been adapted from \citet{doury2024suitability} and is an asymmetric U-Net with a longer expansion path and a one-dimensional bottleneck that accepts the per-timestep normalisation statistics and external forcings as auxiliary inputs, allowing the model to retain large-scale information. An asymmetric precipitation loss penalises under-prediction of heavy events more strongly than over-prediction. A standard MSE loss is used for temperature. 

\item \textbf{Prithvi-UNet [JPL]} fine-tunes NASA's Prithvi-WxC geoscience foundation model with a convolutional U-Net decoder, fine-tuning only selected layers, requiring only 10 epochs per experiment. Coarse predictors are bilinearly interpolated to the high-resolution grid prior to input, and both inputs and targets are normalised using pre-computed statistics. This is the only foundation model algorithm in the benchmark \citep{schmude2024prithvi}. 

\item \textbf{GNN4CD [ICTP]} is a Graph Neural Network (GNN) combining a Gated Recurrent Unit (GRU)-based temporal encoder with graph convolution and graph attention layers. This updated implementation is adapted from the model described in \citet{blasone2025graph}. It employs a processor including three residual Graph Attention Network Convolution blocks (GATv2Conv) with additive aggregation and Layer normalisation. For precipitation, a hybrid loss combining Mean Squared Error (MSE), quantile-aware MSE (QMSE), and Power Spectral Density (PSD) constraints is used to better represent extremes and spatial structure, while temperature is modelled using Gaussian negative log-likelihood to predict both mean and uncertainty. This is the only GNN model in the benchmark, flexible to operate directly on the irregular spatial structure of climate data, without interpolating to a regular grid.

\item \textbf{ParamUNET [LSCE-IPSL]} is a compact U-Net \citep{ronneberger2015u} trained with negative log-likelihood to output the parameters of a Bernoulli-Gamma distribution for precipitation and a Gaussian distribution for temperature, with the expected value submitted as the deterministic point estimate \citep[e.g., ][]{cannon2008probabilistic}. This model is included in the benchmark since it is the backbone for the ParamDiffusion generative model, and has been adapted from \citet{legasa_regional_2026}. 

\item \textbf{DeepESD-IFCAv1 [IFCA]} is a deep CNN comprising three convolutional layers (50, 25, and 1 kernels) followed by a dense output layer, using an asymmetric loss for precipitation and MSE for temperature. In the orography variant, high-resolution orography is appended after the final convolutional layer. Training runs 200--600 epochs, with the number of training epochs determined by early stopping on the validation loss \citep{bano2022downscaling, gonzalez2025deep}.

\item \textbf{DeepESD-IDL [IDL]} shares the CNN architecture of DeepESD-IFCAv1, but replaces the loss function with a Bernoulli-Gamma negative log-likelihood for precipitation and MSE for temperature, where the different loss components for temperature and precipitation were added together with equal weighting \citep{soares2024high}. 

\item \textbf{DeepSensor [BAS]} is a neural process model \citep{andersson2023environmental} that applies convolutional filters to predict Gaussian distribution parameters over the target domain. Neural processes are inherently probabilistic, providing principled estimates of predictive uncertainty, however for this benchmark, deterministic predictions are obtained as the posterior mean. Training uses early stopping, typically converging within 150--180 epochs.

\item \textbf{ANN [BSC]} is a shallow multi-layer perceptron (two layers with 25 and 15 neurons). A Bernoulli-Gamma loss is used for precipitation (where the model predicts parameters of the distribution, and the prediction is the expected value) and MSE for temperature, with training duration determined by early stopping. It is the only algorithm trained entirely on CPU hardware \citep{olmo2022statistical}.

\item \textbf{XGBoost [IDL]} provides a gradient-boosted decision tree baseline adapted from \citet{bushenkova2024towards}, trained with RMSE loss for both variables. Inputs are standardised at the grid-box level; key hyperparameters are a maximum tree depth of 6, a learning rate of 0.05, and early stopping after 20 rounds without improvement. 

\end{itemize}

\subsubsection{Generative and Probabilistic Models}

The generative algorithms include GANs, flow matching, and diffusion model frameworks, thereby representing the current state of the art in generative modeling. All produce stochastic ensemble outputs that capture uncertainty and spatial variability beyond the reach of deterministic regression. 

\begin{itemize}
\item \textbf{FlowMatching-v1 [Met Office UK]} is a flow matching model built on a U-Net backbone (128 base channels) trained to transport samples from a Gaussian source to the target distribution, via a linear scheduler \citep{wetherell2026flow}. Precipitation is normalised with a log1p transform followed by z-score standardisation. 

\item \textbf{ResGAN and ResGAN-v2 [Earth Sciences NZ]} implement a residual Wasserstein GAN following \citet{rampal2025reliable}, pairing a U-Net ($\sim$3M parameters; DetUNet) that first predicts the conditional mean, with a GAN ($\sim$3M parameters) that then predicts the stochastic residual. A pooled maximum intensity constraint (MSE weighted 4.25$\times$) is incorporated into the loss function to better capture precipitation extremes as in \citet{rampal2025reliable}. The two variants differ in learning rate and activation function: ResGAN-v2 uses LeakyReLU(0.5) for temperature. A similar ResGAN variant has been evaluated extensively over New Zealand in previous work, with a focus on extrapolation, transferability, and extremes \citep{rampal2025reliable, rampal2024extrapolation, rampal2025downscaling, ward2025intercomparison}.

\item \textbf{RCMFlow [ESNZ]} is a flow matching model adapted from the \citet{lipman2022flow} implementation. It uses a ResGAN-style architecture (including activation functions) with additional FiLM layers to embed the flow time variable and enable interaction with the predictor fields \citep{ward2025intercomparison}. The model is non-residual and learns the full conditional distribution directly, rather than applying a residual correction, using $\sim$4.5M parameters. Inference uses an Adams–Bashforth third-order ODE solver with 25 flow-time steps per sample for efficiency. Training uses an exponential moving average of model weights to promote stable convergence, with learning rate decay, and is run for a fixed 500 epochs per experiment.

\item \textbf{SpaGAN [KIT]} uses a U-Net2D generator with a convolutional discriminator, trained using a composite loss combining L1, MSE, adversarial, and diversity terms. Sinusoidal day-of-year encoding provides temporal conditioning, and a stochastic noise channel appended at each forward pass generates five ensemble members per time step. Precipitation is log10-transformed and then scaled to the range [-1, 1]. The model is related to that of \citet{glawion2023spategan,glawion2025global}, who used a similar non-residual GAN for downscaling coarse precipitation inputs. The number of training epochs is determined by early stopping on the validation loss.

\item \textbf{EnScale / EnScale-linex [ETH Zurich]} implements a multi-step downscaling framework that first maps atmospheric predictors to spatially pooled targets, then iteratively enhances resolution through sparse localised layers, trained with a proper scoring rule as a loss function \citep{schillinger2025enscale}. EnScale-linex extends this by fitting a linear model first and applying the non-linear EnScale to the residuals. For the ESD setting, an additional stationarity assumption for the residuals is enforced by detrending predictors for EnScale. 

\item \textbf{UiBCorrDiff [UiB]} is a corrective Elucidated Diffusion Model (EDM) comprising a regression U-Net trained with Tweedie deviance loss (p = 1.6) for precipitation and MSE for temperature, followed by a residual EDM diffusion U-Net trained with score matching \citep{mardani2025residual}. The approach was initially developed to downscale from ERA5 to convection permitting scales, but adapted here for downscaling in a climate context. Both modules use a 128-channel, 5-block U-Net architecture. Precipitation is scaled by the global 99th percentile of training-period values; ensemble members are generated via different random seeds passed to a deterministic 36-step sampler. Inference uses an 18-step second-order stochastic sampler. No optimal checkpoint selection strategy was employed. 

\item \textbf{CorrDiff-TW1 [NTNU/CWA]} uses the same EDM-based corrective diffusion architecture as UiBCorrDiff, implemented in NVIDIA PhysicsNeMo (formerly Modulus), comprising a regression U-Net followed by a residual EDM diffusion U-Net. Both modules use a 128-channel U-Net with 5 resolution levels. Unlike UiBCorrDiff, the regression network is trained with an MSE loss for both precipitation and temperature, without a Tweedie deviance loss. Both precipitation and temperature targets are standardized by per-channel z-score normalization over the training period, following the default CorrDiff preprocessing rather than applying a specific precipitation transform. This variant is trained for substantially more epochs than the UiBCorrDiff implementation.

\item \textbf{ParamDiffusion-orog [LSCE-IPSL]} is a diffusion model that uses ParamUNET as its parametric background (predictive information). Precipitation is log-transformed and standardised prior to training, with separate models per variable. The model uses orographic predictors. A single model has been trained for all regions, as opposed to training a separate model for each region. A comprehensive overview of the model can be found in \citet{legasa_regional_2026}.

\item \textbf{RCMGEM-mv-orog [UoB]} is a score-based diffusion model using the NCSN++ architecture under a sub-Variance-Preserving SDE formulation \citep{song2020score}, trained with the Adam optimiser at a learning rate of $2\times10^{-4}$. The number of training epochs varies substantially by domain and experiment (260--2,000 epochs), with the optimal checkpoint selected using validation metrics including spectral density and seasonal bias. The architecture is the same as that of the Convection-Permitting Model Generative Emulator (CPMGEM) of \citet{addison2026machine}, here applied to an RCM setting, adapted to produce multi-variate output and include orography as an input variable.

\item \textbf{ViT-IFCAv1 [IFCA]} is a Vision Transformer \citep{dosovitskiy2020image} in which predictors are partitioned into patches, embedded, and passed through Transformer blocks before upscaling to the target resolution. Stochasticity is injected via FiLM. A CRPS-Spectral loss jointly penalises errors in spatial and frequency domains \citep{nordhagen2025high}, and training duration is determined by early stopping on a held-out validation split. 

\end{itemize}

\subsubsection{Training Data normalisation}

The benchmark imposed no prescribed normalisation strategy, allowing groups to adopt their own approaches. This was done deliberately to enable the community to explore how these decisions interact with architecture and loss function design. The strategies adopted across approaches show both broad commonalities and instructive differences, which are further detailed in Supplementary Table T2.

\textbf{Predictor normalisation}. The majority of approaches standardised input predictors to zero mean and unit variance on a per-channel basis over the full training period
\begin{equation}
    X'(t,c,x,y) = \frac{X(t,c,x,y) - \bar{\mu}(c)}{\sigma(c)},
\end{equation}
where $\bar{\mu}(c)$ and $\sigma(c)$ are the mean and standard deviation of channel \textit{c} computed across all time steps and spatial locations in the training period. Three notable departures from this are worth noting. First, ViT-IFCAv1, DeepESD-IFCAv1 and DeepESD-IDL are standardised at the individual grid-point level rather than globally across space, preserving local climatological gradients in the normalised inputs. Second, Rossby-UNet normalised predictors relative to the local monthly climatology rather than the full training-period mean, so that the normalised anomaly is referenced to the seasonal cycle rather than the annual mean. Third, CNRM-UNeT applied per-timestep normalisation,
\begin{equation}
    X'(t,c,x,y) = \frac{X(t,c,x,y) - \bar{\mu}(t,c)}{\sigma(t,c)},
\end{equation}
where $\mu(t,c)$ and $\sigma(t,c)$ are the spatial mean and standard deviation of channel \textit{c} at time step $t$ only. While this approach removes the instantaneous large-scale state from each input field, the models are also fed the instantaneous normalisation statistics --- the per-timestep mean and standard deviation --- as additional inputs to the network. Lastly, SpaGAN and DeepSensor apply a min-max normalisation to inputs rather than standardisation. Orography, the only static predictor, was normalised separately in most cases using global min-max scaling to [0, 1].

\textbf{Target normalisation}. For the target variable precipitation, a wide range of different normalisation approaches has been used. Most models applied a logarithmic transform prior to training with several subsequently applying z-score standardisation to the transformed values (FlowMatching-v1, ResGAN, DetUNet, ParamDiffusion-orog, Prithvi-UNet, RCMFlow). Additionally, UiBCorrDiff scaled precipitation by the global 99th percentile of training-period gridpoint values, and SpaGAN applied a log10 transform followed by rescaling to [-1, 1]; EnScale normalised by the per-location standard deviation; CorrDiff-TW1 applied  z-score normalization to precipitation and RCMGEM-mv-orog applied a square-root transformation followed by normalisation to [-1, 1]. Also, ParamUNET, ANN and DeepESD-IDL did not explicitly apply a normalisation because they bypassed explicit transformations by directly predicting the parameters of a Bernoulli-Gamma distribution \citep[e.g.,][]{cannon2008probabilistic, bano2020configuration, rampal2022high}. In contrast, some other models applied no explicit normalisation to precipitation. For temperature, normalisation was consistent across different ML approaches, with the large majority applying grid-point-level z-score standardisation computed over the training period and reapplied at inference. A small number of models instead rescaled temperature to [-1, 1] or [0, 1] to match the output range of their architecture. Additionally, a few models applied no normalization at all to this variable, working directly with raw temperature values.

\subsection{Evaluation Metrics}
\label{subsection:evaluation-metrics}

The core evaluation metrics are summarised in Table~\ref{tab:metrics}; unless stated otherwise, all are computed at the grid-point level and spatially averaged over each domain. Each metric is briefly introduced below.

\begin{table}[ht]
\centering
\renewcommand{\arraystretch}{1.4}
\caption{Core evaluation metrics used in CORDEX-ML-Bench, applied to each of the
three pilot domains (ALPS, NZ, SA) and as a domain average (AV). Smaller absolute
values are better for all metrics (lower is better) except for Perkins Skill Score. Metrics
are computed over the historical (1981--2000) and mid-century future (2041--2060)
test periods.}
\label{tab:metrics}
\begin{tabular}{@{}llp{7.2cm}@{}}
\toprule
\rowcolor{lightgray}
\textbf{Metric} & \textbf{Variable} & \textbf{Description} \\
\midrule
RMSE                   & pr, tasmax & Root-mean-square error (RMSE) of daily values \\ \hline
Clim Mean              & pr, tasmax &  RMSE of the climatological
                                       mean \\ \hline
SDII                   & pr         &  RMSE of the
                                       Simple Daily Intensity Index (SDII; mean
                                       precipitation on wet days,
                                       $\geq$1\,mm\,day$^{-1}$) \\ \hline
Rx1day                 & pr         & RMSE of the climatological
                                       mean annual maximum 1-day precipitation \\ \hline
TXx                    & tasmax     & RMSE of the climatological
                                       mean annual maximum daily maximum
                                       temperature \\ \hline
RALSD                  & pr, tasmax & Radially averaged log-spectral distance;
                                       measures spatial power-spectral variability \\ \hline
LHD                    & pr         & Logarithmic histogram distance; measures
                                       distributional fit \\ \hline
PSS                    & tasmax     & Perkins Skill Score; measures distributional
                                       overlap \\ \hline
$\Delta$M signal error & pr, tasmax & Spatial RMSE of the climate-change
                                       signal (future $-$ historical climatological
                                       mean) expressed as a percentage \\
\bottomrule
\end{tabular}
\end{table}
\begin{itemize}


\item \textbf{Root-mean-square error (RMSE)}.

The primary measure of day-to-day skill used in this study is RMSE, defined as
\begin{linenomath*}
\begin{equation}
\mathrm{RMSE}=\sqrt{\frac{1}{N_tN_xN_y}\sum_{t,x,y}\left(\hat{Y}_{t,x,y}-Y_{t,x,y}\right)^2},
\end{equation}
\end{linenomath*}
where $\hat{Y}_{t,x,y}$ and $Y_{t,x,y}$ denote the predicted and reference (ground truth RCM) fields at
time step $t$ and grid point $(x, y)$, and $N_t$, $N_x$, and $N_y$ are the numbers of
time steps and grid points in each dimension. RMSE is computed separately for daily
precipitation (pr) and daily maximum temperature (tasmax).

\vspace{0.15in}

\item \textbf{Simple Daily Intensity Index (SDII)}: The SDII, developed by the Expert
Team on Climate Change Detection and Indices \citep[ETCCDI;][]{zhang2011indices}, measures mean
precipitation intensity on wet days (days with pr $\geq 1$\,mm\,day$^{-1}$),
\begin{linenomath*}
\begin{equation}
\mathrm{SDII}=\frac{\sum_{t:\,pr_t\geq 1} pr_t}
                   {\sum_{t:\,pr_t\geq 1} 1},
\end{equation}
\end{linenomath*}

where the sums run only over wet days ($pr_t \geq 1$\,mm\,day$^{-1}$). SDII captures
whether a model correctly partitions total rainfall between wet-day frequency and
wet-day intensity. We evaluate it by computing the RMSE of the SDII climatology (i.e. over a 20-year period) from
the predicted field against that from the reference simulation over the period of
interest.
\vspace{0.1in}

\item \textbf{Rx1day and TXx}: Skill in reproducing extremes is quantified using two ETCCDI indices: the annual maximum one-day precipitation (Rx1day) and the annual maximum daily maximum temperature (TXx). Both are computed at each grid point, averaged over all years in the evaluation period, and scored as the RMSE between the predicted and reference climatology.
\vspace{0.1in}

\item \textbf{Interannual Variability (IAV)}: For maximum temperature we additionally report the interannual variability of annual means. The standard deviation is computed from annual means across time at each grid point, and the RMSE is then calculated between the predicted and true interannual variability. This provides a measure of temporal variability and assesses how well the models reproduce RCM simulation year-to-year fluctuations/variability.

\vspace{0.1in}
\item \textbf{Radially-averaged Log Spectral Distance (RALSD)}. A well-known limitation of regression-based downscaling is the tendency to produce overly smooth spatial fields, as predictions regress toward the mean and fine-scale variability is underestimated \citep{rampal2025reliable, subich2025fixing, ravuri2021skilful}. This metric evaluates how accurately a model reproduces spatial variability within precipitation or temperature fields across synoptic to fine scales. Following \citet{harris2022generative, rampal2025reliable}, we compute the radially averaged two-dimensional power spectral density (PSD) for both predicted ($S_{\hat{Y}}(k)$) and reference ground truth fields ($S_Y(k)$) and subsequently calculate the log-spectral distance:
\begin{linenomath*}
\begin{equation}
\mathrm{RALSD} (dB)=\sqrt{\frac{1}{K}\sum_k \left[\log S_Y(k)-\log S_{\hat{Y}}(k)\right]^2}.
\end{equation}
\end{linenomath*}
where $k$ indexes the radial wavenumber and $K$ is the total number of wavenumber bins. RALSD is computed for each day and then averaged across all days. As with most of the metrics, the smaller the value, the better.
\vspace{0.1in}

\item \textbf{Perkins Skill Score (PSS)}: The PSS \citep{perkins2007evaluation} quantifies the overlap between the predicted and reference probability density functions as the sum of the minimum counts across all histogram bins:
\begin{linenomath*}
\begin{equation}
\mathrm{PSS} = \sum_{b=1}^{B} \min\!\left(H_Y(b),\, H_{\hat{Y}}(b)\right),
\end{equation}
\end{linenomath*}
where $H_Y(b)$ and $H_{\hat{Y}}(b)$ are the normalised histogram counts for the reference and predicted distributions in bin $b$, respectively, and $B$ is the total number of bins. PSS ranges from 0 (no overlap) to 1 (perfect agreement). It is applied here to the distribution of daily maximum temperature across all times and locations, rather than on a per-location basis for simplicity. An example illustrating the calculation and interpretation of the PSS for model evaluation is provided in Figure S1.

\vspace{0.1in}

\item \textbf{Logarithmic Histogram Distance (LHD)}: quantifies how well a model reproduces the full distribution of a variable, analogous to RALSD in spectral space. Contrary to PSS which weights all bins equally and is therefore more sensitive to errors near the mode of the distribution, LHD applies a logarithmic weighting in histogram space so that errors across all intensity bins contribute equally, giving greater sensitivity to the tails of heavy-tailed distributions (e.g.\ precipitation). It is computed as the root-mean-square distance between the logarithms of normalised frequency histograms of the predicted ($\hat{Y}$) and reference ($Y$) fields:
\begin{linenomath*}
\begin{equation}
    \mathrm{LHD} (dB) = \sqrt{\frac{1}{B}\sum_{b=1}^{B} \left[\log H_Y(b) - \log H_{\hat{Y}}(b)\right]^2},
\end{equation}
\end{linenomath*}
where $H_Y(b)$ and $H_{\hat{Y}}(b)$ are the normalised histogram counts in bin $b$ and $B$ is the total number of bins. Lower values indicate a closer match to the reference distribution. The metric is only computed for bins where both distributions contain more than 10 counts, and histograms are computed over all times and locations, as with PSS. Figure S2 provides an example illustrating the calculation and interpretation of the LHD for model evaluation.

\vspace{0.1in}
\item \textbf{Climate-change signal}: The climate-change signal error ($\epsilon_{\Delta M}$) measures how well a model captures the change in a climatological metric \textit{M} between a historical period (1981--2000) and a mid-century period (2041--2060):
\begin{linenomath*}
\begin{equation}
\epsilon_{\Delta M}=\Delta \hat{M}-\Delta M,\qquad \Delta M=M_{future}-M_{historical}.
\end{equation}
\end{linenomath*}
For temperature metrics, $\Delta M$ is an absolute change; for precipitation metrics, it is a relative change ($\Delta M$ / $M_{historical}$). The climatological metric \textit{M} may represent changes in mean fields, such as temperature, or in extremes, such as Rx1day. This error reveals whether a model faithfully extrapolates beyond its training distribution (ESD) or interpolates across historical and future conditions (RCM Emulator), and is therefore treated as one of the most important fitness-for-purpose criteria, though its interpretation requires care. Here, we provide an initial assessment of the models' ability to reproduce future changes in extremes for an unseen period within the training GCM, with further analysis across seen and unseen GCMs and the imperfect evaluation setting deferred to forthcoming studies.

\end{itemize}

For generative models, predictions are produced as ensembles of 5 or 10 members, depending on what each contributor to the benchmark was able to generate. For diagnostics targeting mean aspects of the climate we use the ensemble mean. For extreme metrics, the statistic is first computed for each ensemble member, then averaged across members before the final metric/score is calculated. For metrics such as RALSD, PSS, LHD, the score is computed for each member, before averaging the score across all members. This approach provides a consistent basis for evaluating the added value of generative approaches.

\subsection{Reporting and scorecard format}

All metrics described above are reported for both the ESD and RCM Emulator experiments and are summarised in a scorecard. The historical period (1981--2000) is excluded from training in both experiments; therefore, evaluation over this held-out period represents perfect cross-validation, shown in the upper triangle of the scorecard (Figures \ref{fig:score-pr}-\ref{fig:score-tasmax}). For the future period (2041--2060), evaluation represents either perfect extrapolation when models are trained only on historical data (ESD), or perfect interpolation in the RCM Emulator configuration, where models are trained on combined historical and future data. These scores are shown in the lower triangle of the scorecard. The dual-panel scorecard format introduced in Section~\ref{sec:benchmarking} enables direct comparison of in-sample and out-of-sample skill within each experiment, while also facilitating comparison between the two experiments.

\section{Results}

In Section \ref{sec:case_studies}, we first present several case studies to qualitatively illustrate model skill, followed by skill in predicting climatologies in \ref{sec:climatology}. Both use the RCM Emulator experiment over the cross-validation period (1981--2000), which provides a representative overview of results across both experiments. Section~\ref{sec:cc_signal} then examines climate change signal skill for both the ESD and RCM Emulator experiments over the mid-century period (2041-2060) -- a period withheld from training. Finally, Section~\ref{sec:benchmarking} provides a comprehensive multi-metric benchmarking and ranking across historical and future periods. Throughout, evaluation is restricted to the training GCM in an unseen out-of-sample period under the perfect predictor framework, with a more in-depth analysis of GCM transferability (imperfect evaluation) left for future studies.

\subsection{Case Studies}\label{sec:case_studies}

\begin{figure}
\noindent\includegraphics[width=0.9\textwidth, trim=0 0 0 25mm, clip]{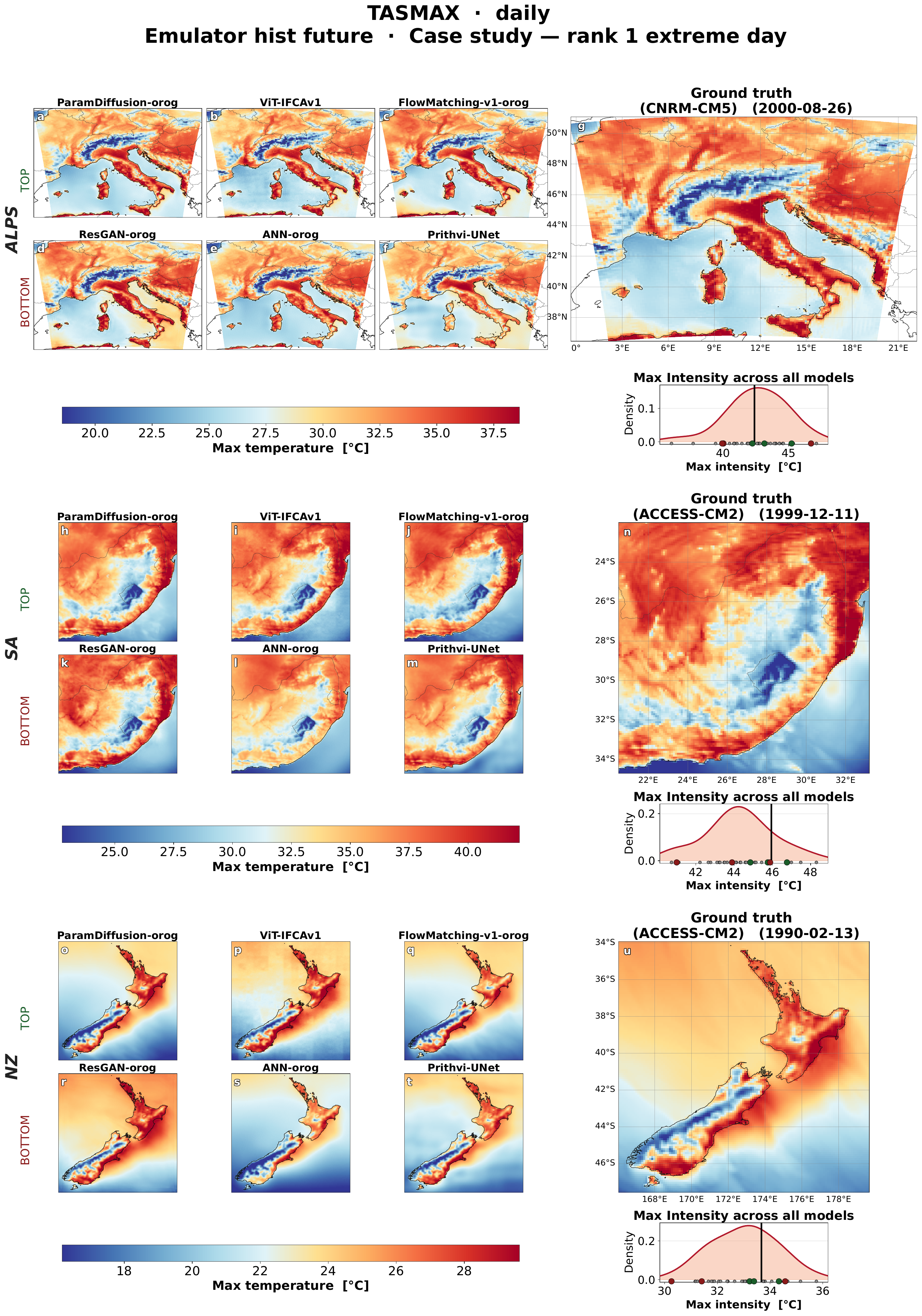}
\caption{An extreme-event case study is included to illustrate predictions from different ML models in the RCM Emulator experiment across all three domains. The selected event (2000-08-26 for the European Alps, 1999-12-11 for South Africa, and 1990-02-13 for New Zealand) corresponds to the hottest day from the cross-validation historical period (1981--2010), based on area-averaged daily maximum temperature, representing a widespread extreme heat event in each domain. Six models are shown per domain: three drawn from the top-10 ranked models (ParamDiffusion-orog, VIT-IFCAv1, FlowMatching-v1-orog; green labels) and three from the bottom 20 (ResGAN-orog, ANN-orog and Prithvi-UNet; red labels). For the generative models, only the first ensemble member is show here. The inset beneath the ground-truth panel shows the distribution of spatial maximum intensity across all 40 models for that day, with the vertical black line indicating the observed value. Fields are displayed on the native $\sim$10 km target grid.}
\label{fig:case-tasmax}
\end{figure}

\begin{figure}
\noindent\includegraphics[width=0.9\textwidth, trim=0 0 0 25mm, clip]{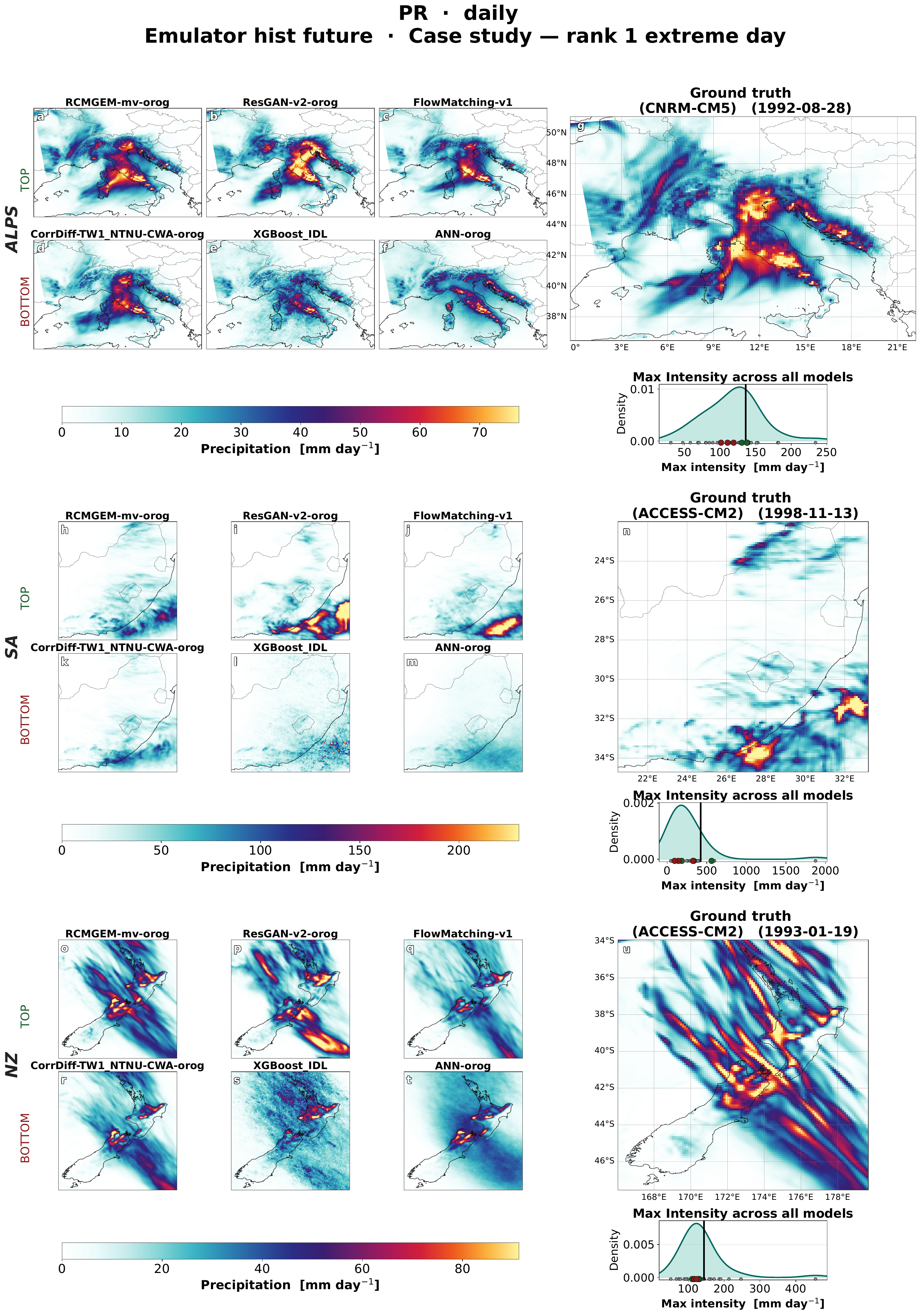}

\caption{Same as Figure~\ref{fig:case-tasmax}, but for the wettest day for precipitation (based on area-averages). In this case the selected events are 1992-08-28 for the European Alps, 1998-11-13 for South Africa, and 1993-01-19 for New Zealand. The three models shown from the top-10 ranked models are RCMGEM-mv-orog, ResGAN-v2-orog, and FlowMatching-v1, while the three models shown from the bottom 20 are CorrDiff-TW1$\_$NTNU-CWA-orog, XGBoost$\_$IDL, and ANN-orog.}
\label{fig:case-pr}
\end{figure}

We begin with case studies across each of the three domains to illustrate how different model architectures predict individual extreme events (Figures~\ref{fig:case-tasmax} and~\ref{fig:case-pr}). These case studies are selected by their domain-averaged intensity: the hottest day for maximum temperature and the wettest day for precipitation, ensuring the illustrated events correspond to widespread rather than localised extremes. The six models shown comprise three drawn from the top-10 and three from the bottom-20 of the overall ranking for each variable, based on the multi-metric scoring described in Section~\ref{sec:benchmarking}. It should be noted that lower-ranked models are not necessarily poor performers in an absolute sense since many of these models have demonstrated skill relative to many conventional statistical downscaling approaches \citep{bano2020configuration, gonzalez2025deep}. Also note that precipitation and temperature have variability that is not entirely predictable using coarse-resolution predictors, with downscaling being an under-constrained problem (ill-posed), so the downscaling methods are not expected to reproduce the high-resolution targets in full detail.

For the maximum temperature case study (Figure~\ref{fig:case-tasmax}), most models reproduce the large-scale spatial patterns well across all three domains, including the cooler temperatures over the European Alps and New Zealand's Southern Alps, the anomalously warm conditions over north-eastern South Africa, and the localised heating over south-eastern New Zealand consistent with the föhn effect. As for the extreme precipitation case study (Figure~\ref{fig:case-pr}), the selected generative models generally produce plausible spatial patterns across all three domains. Over the Alps, RCMGEM-mv-orog, ResGAN-v2-orog, FlowMatching-v1, and CorrDiff capture the orographic enhancement of precipitation over mountainous regions, even when the precise location of maximum intensity is not perfectly predicted. Similar behaviour is evident over New Zealand, where generative approaches reproduce key characteristics of narrow frontal precipitation bands extending from the north-west to south-east, which generate strong orographic precipitation over north-eastern parts of the South Island. Over Southern Africa, these generative models capture aspects of larger convective structures, consistent with the mesoscale convective organisation typical of austral-summer circulation. In contrast, deterministic regression models — most notably XGBoost\_IDL and ANN-orog — systematically underestimate extreme intensities and produce spatially smooth or noise-dominated fields that lack the fine-scale structure expected of a $\sim$10 km precipitation field, with this limitation most pronounced outside the Alps domain. This behaviour is a well-documented consequence of regression-to-the-mean in ML-based downscaling \citep{rampal2025reliable, ravuri2021skilful, harris2022generative, vosper2023deep} and is assessed more formally through the RALSD metric (Section \ref{sec:benchmarking}). Additional case studies, including those for the ESD experiment, are shown in Figures~S3--S6.


\subsection{Historical Climatologies}\label{sec:climatology}

\begin{figure}
\noindent\includegraphics[width=0.9\textwidth, trim=0 0 0 25mm, clip]{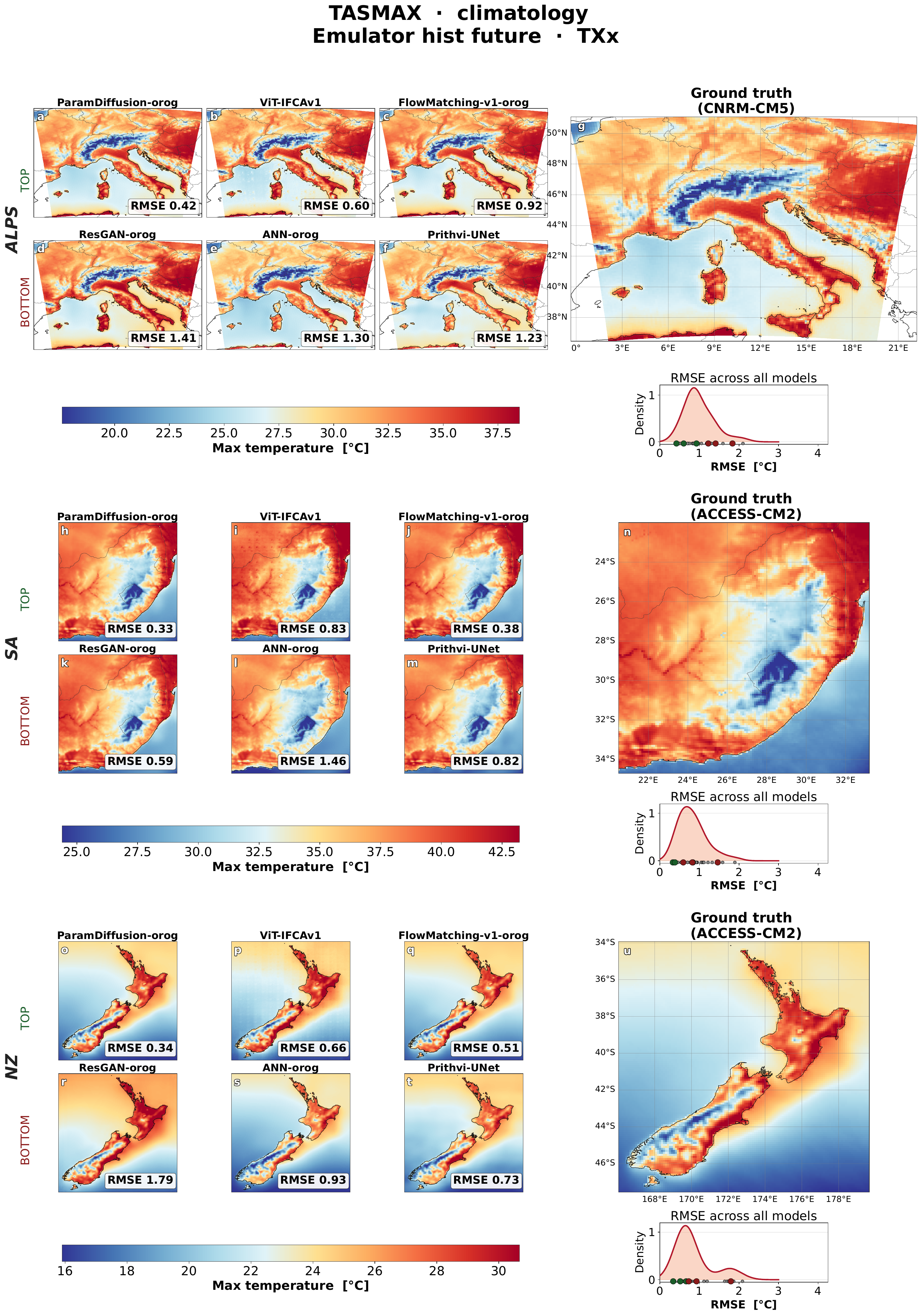}
\caption{TXx climatology (mean annual maximum daily maximum temperature) over the cross-validation period (1981--2000) for the RCM Emulator experiment. Layout as in Figure~\ref{fig:case-tasmax}. The three top-ranked models shown are ParamDiffusion-orog, ViT-IFCAv1, and FlowMatching-v1-orog; the three bottom-ranked models are ResGAN-orog, ANN-orog, and Prithvi-UNet. RMSE relative to the reference RCM climatology is annotated in the bottom-right corner of each panel, and is indicated in units of $^\circ$C. The inset below the ground-truth panel shows the distribution of TXx RMSE across all 40 models; green and red dots indicate the sampled top and bottom models, respectively, and grey dots indicate all other models. Fields are displayed on the native $\sim$10 km target grid.}
\label{fig:TXx-clim}
\end{figure}

\begin{figure}
\noindent\includegraphics[width=0.9\textwidth, trim=0 0 0 25mm, clip]{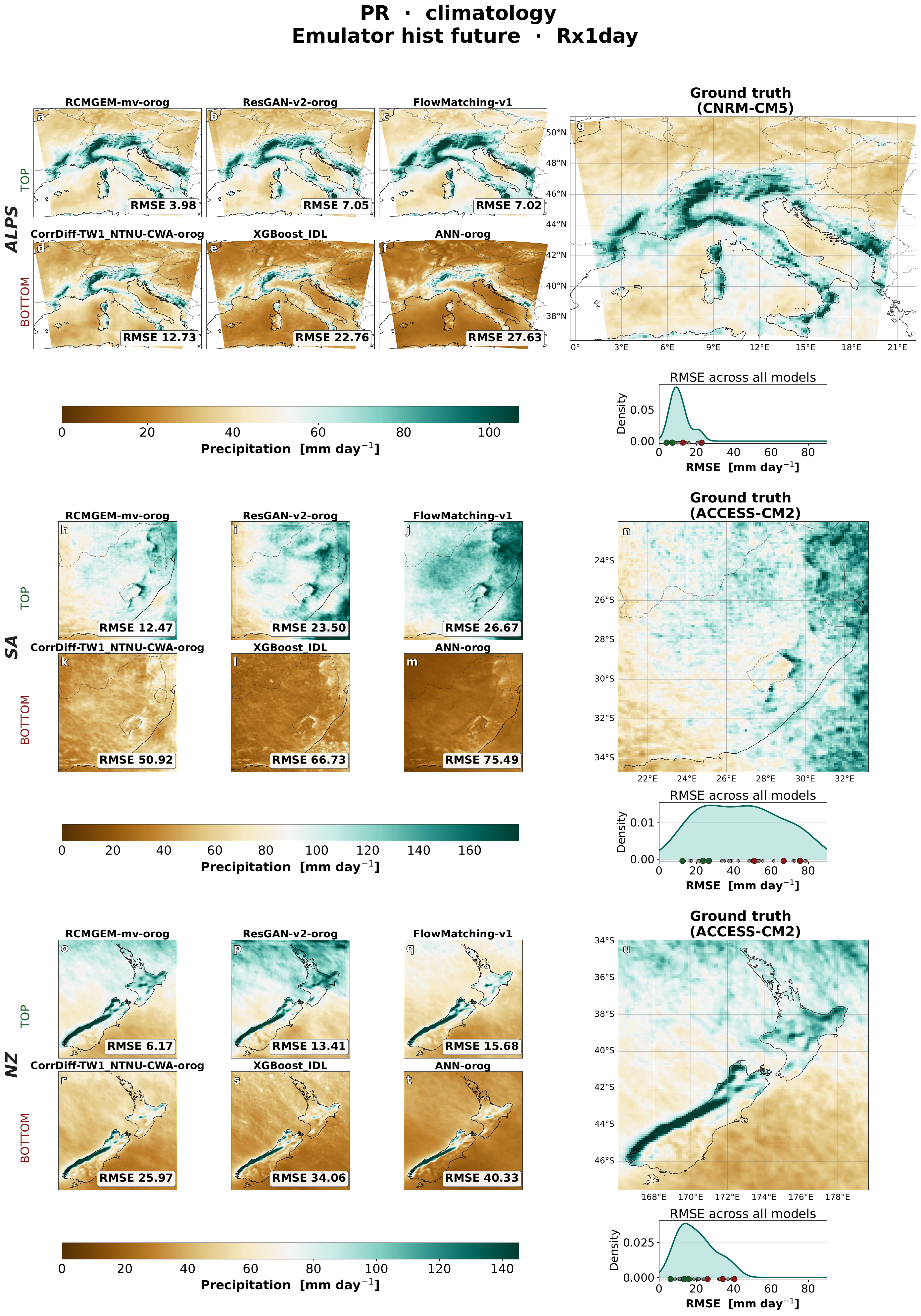}
\caption{Same as Figure~\ref{fig:TXx-clim}, but for the Rx1day climatology (mean annual maximum one-day precipitation). The six model panels show three models drawn from the top-10 ranked configurations (green labels; RCMGEM-mv-orog, ResGAN-v2-orog, FlowMatching-v1) and three drawn from the bottom 20 (red labels; CorrDiff-TW1\_NTNU-CWA-orog, XGBoost\_IDL, ANN-orog). Units of RMSE are in mm day$^{-1}$.}
\label{fig:rx1day-clim}
\end{figure}

We next assess climatological skill using two ETCCDI indices: the annual maximum daily maximum temperature (TXx) and annual maximum one-day precipitation (Rx1day), evaluated over the cross-validation period (1981–2000) for the RCM Emulator experiment (Figures~\ref{fig:TXx-clim} and \ref{fig:rx1day-clim}).

For the TXx climatology (Figure~\ref{fig:TXx-clim}), most models reproduce the spatial climatology faithfully across all three domains, capturing the cooler temperatures over the mountainous regions of the European and New Zealand Alps, land--sea temperature contrasts over the Italian peninsula and New Zealand, and the warm interior of Southern Africa. The leading models in the RCM Emulator experiment --- ParamDiffusion-orog, ViT-IFCAv1, and FlowMatching-v1-orog --- achieve RMSE values of 0.3--0.9\,$^\circ$C, with the majority of the 40 models clustering below 2\,$^\circ$C across all domains. Even the lower-ranked models shown --- ResGAN-orog (RMSE; 0.59--1.79\,$^\circ$C), ANN-orog (0.93--1.46\,$^\circ$C), and Prithvi-UNet (0.73--1.23\,$^\circ$C) --- broadly reproduce the large-scale spatial patterns of TXx, though some exhibit systematic biases; for example, ResGAN-orog captures the spatial structure well but shows a consistent warm bias across all domains. 

For the Rx1day climatology (Figure~\ref{fig:rx1day-clim}), the top-ranked models --- RCMGEM-mv-orog, ResGAN-v2-orog, and FlowMatching-v1 --- reproduce the broad spatial patterns of annual maximum precipitation across all three domains, including orographic enhancement along mountain ranges and the north-south gradient over New Zealand. Bottom-ranked models show substantially larger errors; for example, over Southern Africa, ANN-orog achieves an RMSE of 75\,mm\,day$^{-1}$ compared to 12.5\,mm\,day$^{-1}$ for RCMGEM-mv-orog. A dry bias and systematic underestimation of extreme precipitation is a consistent feature of the bottom-ranked models, including some generative approaches such as those based on CorrDiff. The inter-model spread and typical RMSE values are largest over Southern Africa and smallest over the Alps across the full model ensemble. We attribute the larger RMSE in the South African domain to the dominant role of mesoscale convection during the austral summer \citep{blamey2013role}, which we suspect are somewhat less predictable from the set of large-scale predictor fields used here; though this will be explored in greater depth in a companion paper. It is worth noting that rank order is not always consistent across metrics and domains. A model's overall rank reflects aggregate skill, and strong performance on one metric does not guarantee skill on another, underscoring the importance of the multi-metric ranking framework discussed in Section~\ref{sec:benchmarking}. Similar results for the ESD experiment are shown in Figures~S7 and S8.

Overall, these results highlight a much larger inter-model spread in skill for precipitation than temperature, with generative models showing a clear advantage over deterministic approaches in capturing fine-scale spatial detail and extremes --- a difference far less pronounced for temperature, where most architectures perform well. The inter-model spread in Rx1day RMSE spans nearly an order of magnitude across domains, compared with roughly 2\,$^\circ$C for TXx. Results here are limited to the RCM Emulator cross-validation period; a comprehensive comparison across experiments, future periods, and the full metric suite is presented in Section~\ref{sec:benchmarking}.


\subsection{Future Climate Change Signals}\label{sec:cc_signal}

\begin{figure}
\noindent\includegraphics[width=0.9\textwidth, trim=0 0 0 25mm, clip]{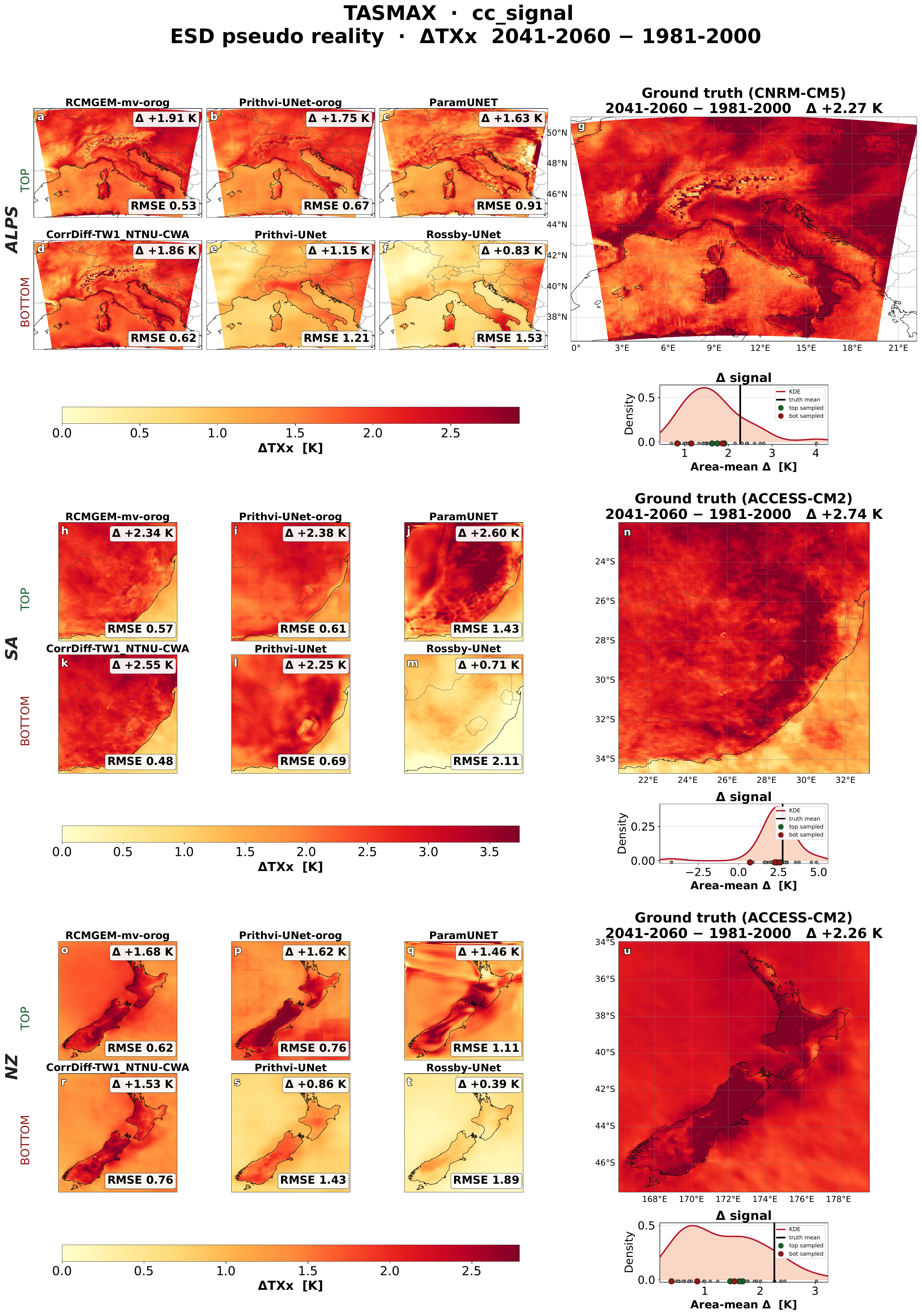}
\caption{Climate-change signal in TXx ($\Delta$TXx, K) for the models trained on the historical period only (ESD experiment), defined as the difference in mean annual maximum daily maximum temperature between the mid-century (2041--2060) and historical (1981--2000) periods. Each row corresponds to one domain (Alps, Southern Africa, New Zealand); the reference signal is shown in the large right-hand panel with the area-mean $\Delta$ annotated. Six models are shown per domain: three from the top of the overall ranking (green labels; RCMGEM-mv-orog, Prithvi-UNet-orog, ParamUNET) and three from the lower end (red labels; CorrDiff-TW1\_NTNU-CWA, Prithvi-UNet, Rossby-UNet). The area-mean $\Delta$ and RMSE relative to the reference change field are annotated per panel. The inset shows the distribution of area-mean $\Delta$TXx across all 40 models; the vertical black line marks the reference value. Reference signals are +2.27 K (Alps), +2.74 K (SA), and +2.26 K (NZ).}
\label{fig:TXx-esd}
\end{figure}

\begin{figure}
\noindent\includegraphics[width=0.9\textwidth, trim=0 0 0 25mm, clip]{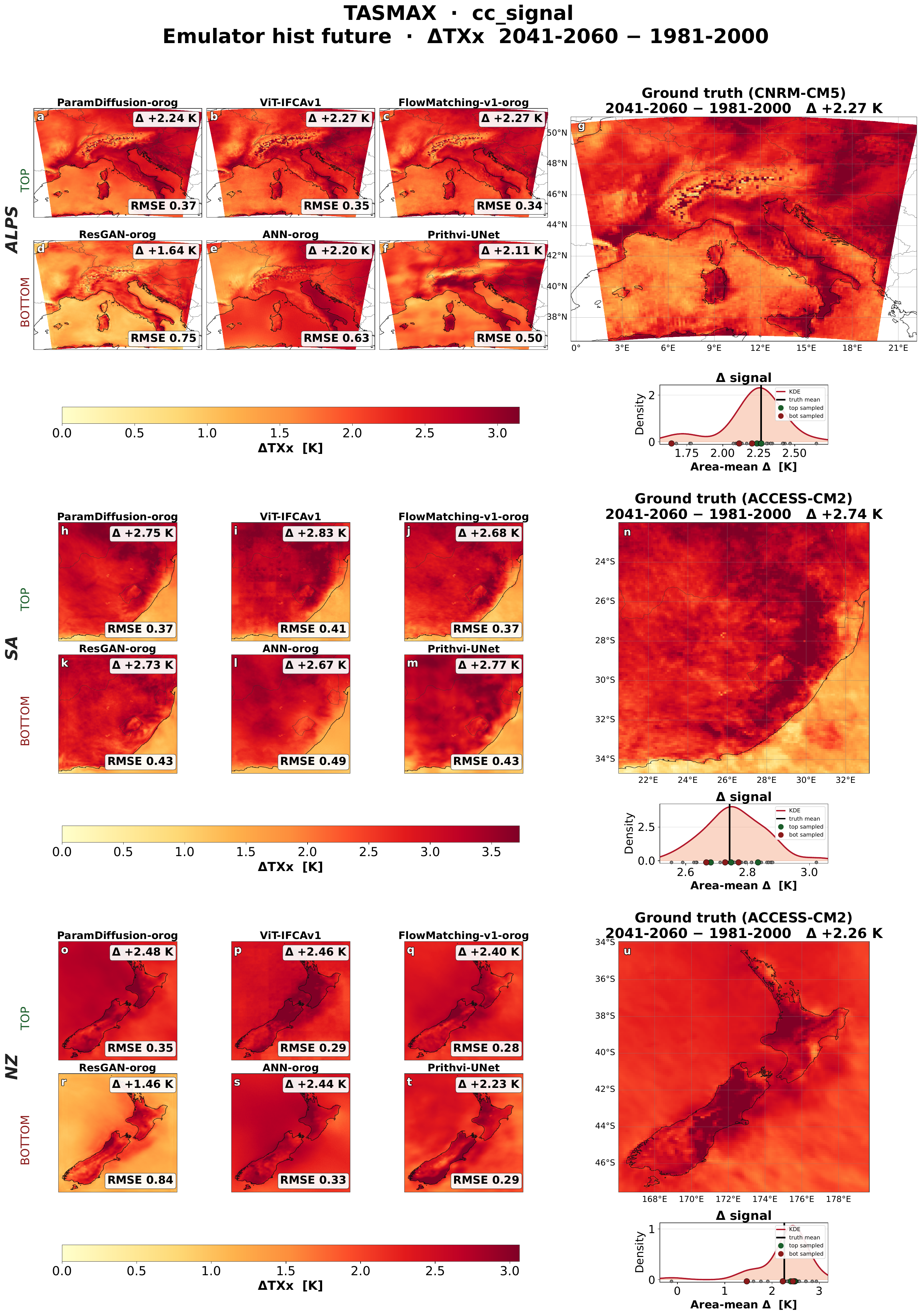}
\caption{As Figure~\ref{fig:TXx-esd}, but for the RCM Emulator experiment. Models shown are ParamDiffusion-orog, ViT-IFCAv1, and FlowMatching-v1-orog (top); ResGAN-orog, ANN-orog, and Prithvi-UNet (bottom). Reference signals are identical to Figure~\ref{fig:TXx-esd}.}
\label{fig:TXx-emul}
\end{figure}

\begin{figure}
\noindent\includegraphics[width=0.9\textwidth, trim=0 0 0 25mm, clip]{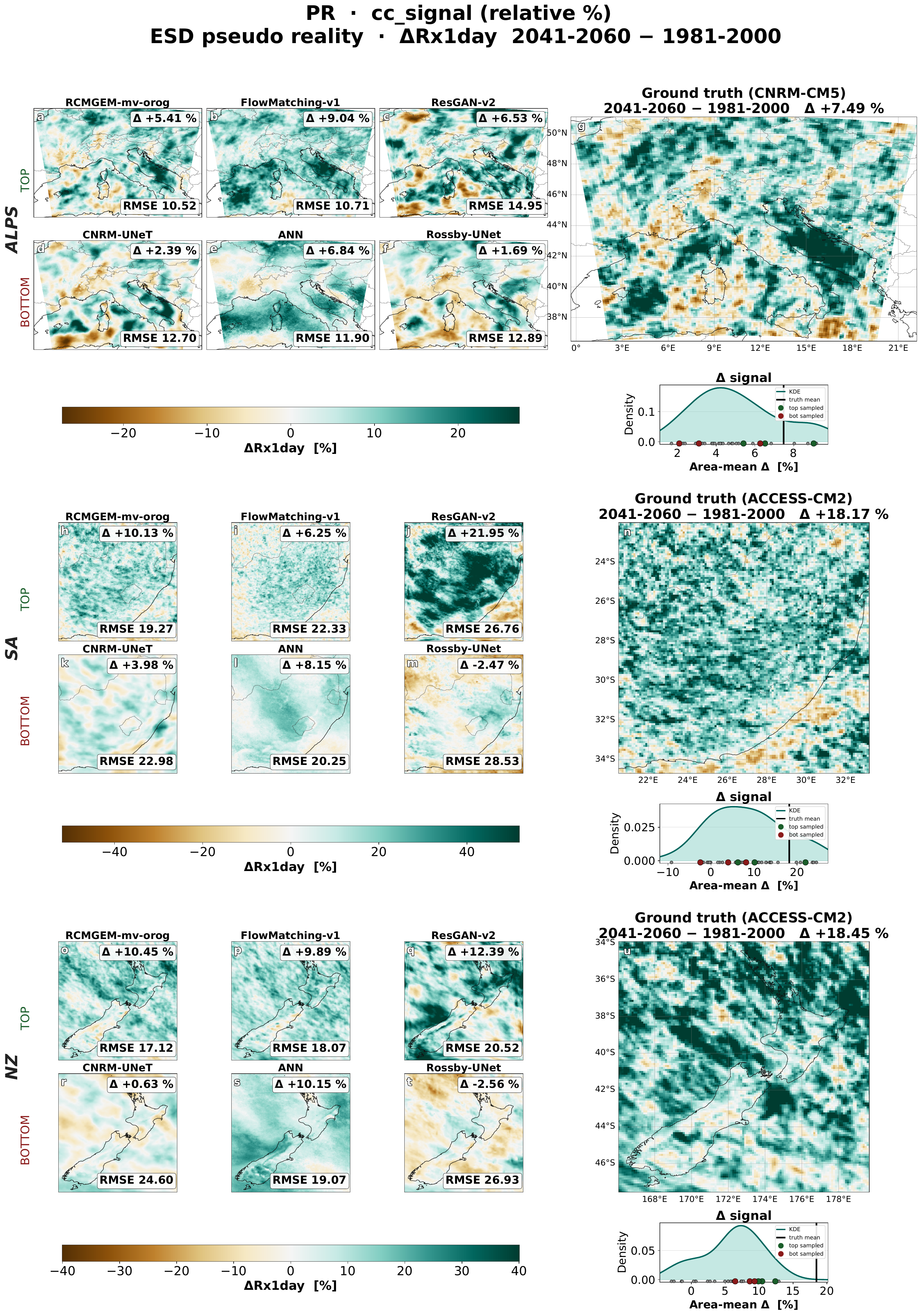}
\caption{Climate-change signal in Rx1day ($\Delta$Rx1day, \%) for the ESD experiment, defined as the relative change in mean annual maximum one-day precipitation between mid-century (2041--2060) and the historical period (1981--2000). Models shown are RCMGEM-mv-orog, FlowMatching-v1, and ResGAN-v2 (top); CNRM-UNeT, ANN, and Rossby-UNet (bottom). The area-mean $\Delta$ and RMSE of the change field are annotated per panel. The inset shows the distribution of area-mean $\Delta$Rx1day across all 40 models; the vertical black line marks the reference domain-mean. Reference signals are +7.49\% (Alps), +18.17\% (SA), and +18.45\% (NZ).}
\label{fig:rx1day-esd}
\end{figure}

\begin{figure}
\noindent\includegraphics[width=0.9\textwidth, trim=0 0 0 25mm, clip]{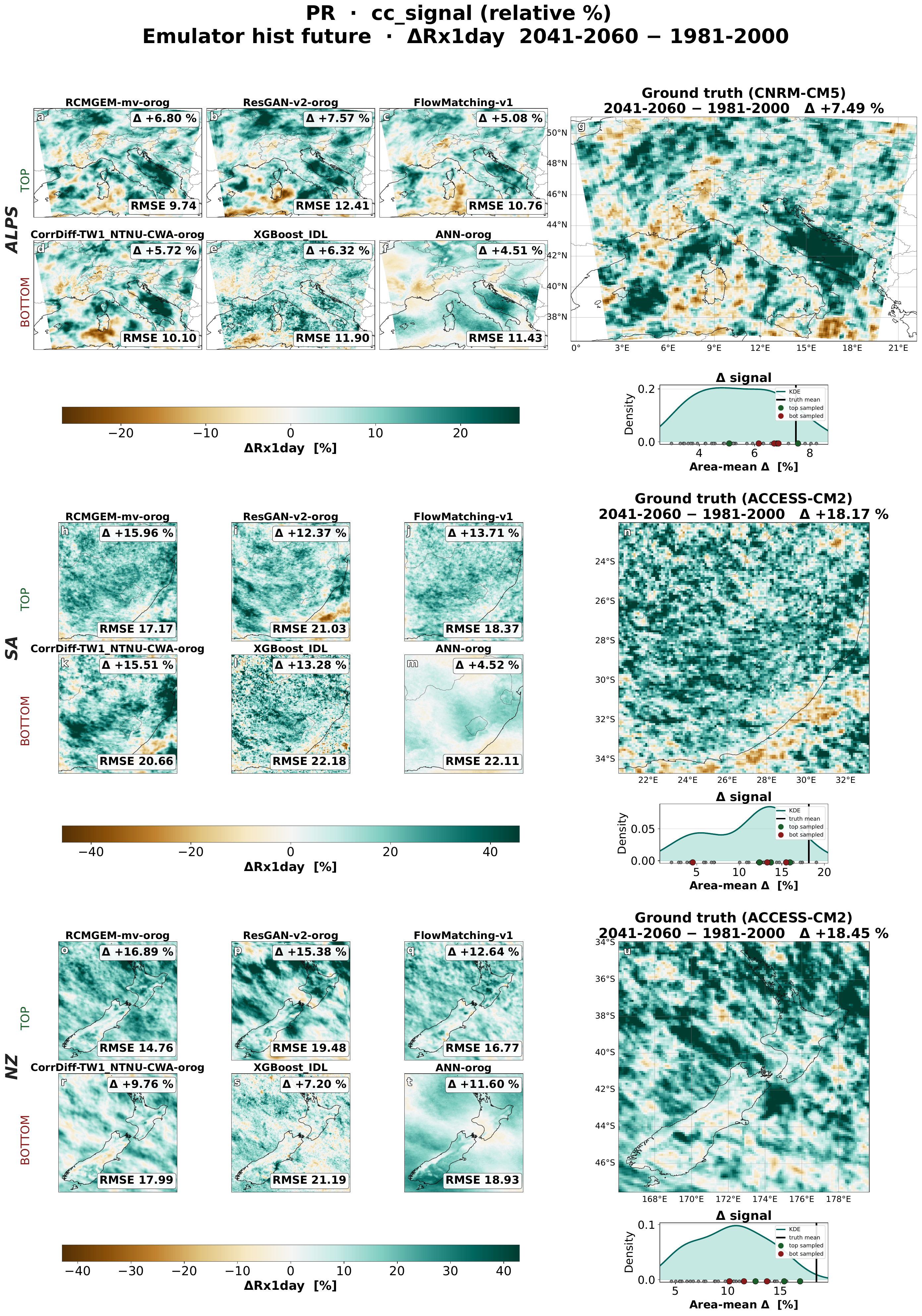}
\caption{As Figure~\ref{fig:rx1day-esd}, but for the RCM Emulator experiment. Models shown are RCMGEM-mv-orog, ResGAN-v2-orog, and FlowMatching-v1 (top); CorrDiff-TW1\_NTNU-CWA-orog, XGBoost\_IDL, and ANN-orog (bottom). Reference signals are identical to Figure~\ref{fig:rx1day-esd}.}
\label{fig:rx1day-emul}
\end{figure}

We now assess the models' ability to reproduce the mid-century climate-change signal (2041--2060 relative to 1981--2000) in TXx and Rx1day --- a period withheld from training in both experiments. Unlike previous sections, we also include results for the ESD experiment, which better tests extrapolation capabilities since models are trained exclusively on historical simulations with no future data involved. Figures~\ref{fig:TXx-esd}--\ref{fig:rx1day-emul} show predicted signals from three top-ranked and three lower-ranked models alongside the reference RCM signal for each domain, with the area-mean $\Delta$ annotated per panel and the full 40-model distribution shown in the bottom-right inset; note that rankings are based on the overall multi-metric score rather than this specific metric. Results for TXx are discussed first (Figures~\ref{fig:TXx-esd}--\ref{fig:TXx-emul}), followed by Rx1day (Figures~\ref{fig:rx1day-esd}--\ref{fig:rx1day-emul}).

The reference (i.e. RCM predicted ground-truth) mid-century warming signal of TXx is +2.27 K over the Alps, +2.74 K over Southern Africa, and +2.26 K over New Zealand. In the ESD experiment (Figure~\ref{fig:TXx-esd}), nearly all 40 models underestimate this signal across all three domains. The distributions of area-mean $\Delta$ are systematically shifted to the left of the reference line, with most models predicting warming of only +1.5--2.0 K --- an underestimation of roughly 0.3--0.8 K depending on domain, though some individual models underestimate this signal even more. This underestimation is evident even among the highest-ranked ESD models --- RCMGEM-mv-orog, Prithvi-UNet-orog, and ParamUNET. While the spatial patterns of warming are broadly plausible, the magnitude of the climate change signal is underestimated for all three domains, and across nearly all models. In contrast, for the RCM Emulator experiment (Figure~\ref{fig:TXx-emul}), the distributions of the mean climate change signal across all models shift toward the reference mean signal, and most models achieve area-mean $\Delta$ values within 0.1--0.3 K of the reference across all domains. The top-ranked models --- ParamDiffusion-orog, ViT-IFCAv1, and FlowMatching-v1-orog --- predict area-mean changes in close agreement with the reference values. The RMSE distributions are also considerably narrower than in the ESD experiment. The main exception is ResGAN-orog, which underestimates the TXx signal in the Alps (+1.63\,K) and New Zealand (+1.43\,K) despite being trained on future data. Overall, these results indicate that including future periods during training is highly beneficial for reproducing future temperature climate change signals, although further analysis of this aspect will be explored in a companion study.

For Rx1day, the reference mid-century change is +7.49\% over the Alps, +18.17\% over Southern Africa, and +18.45\% over New Zealand. Unlike mean precipitation, whose global changes are more constrained by radiative energy balance \citep{trenberth2003changing}, changes in Rx1day are more strongly influenced by increases in precipitable water via Clausius-Clapeyron scaling and are therefore often positive across many regions, though rates of increase vary \citep{pfahl2017understanding, o2015precipitation}. Also, the spatial patterns of this climate-change signal are somewhat noisy, as they are more strongly influenced by individual extreme events not linked to the large-scale conditions represented by the predictors (e.g., convective events). This makes RMSE a less informative metric overall for this diagnostic (Rx1day), making additional qualitative assessment important. In the ESD experiment (Figure~\ref{fig:rx1day-esd}), all models substantially underestimate the area-mean Rx1day climate change signal across all domains, especially for South Africa and New Zealand. The distributions are clustered to the left of the reference, with most models predicting area-mean changes of only +2--8\% against reference values of +7.5--18.5\%. Several models --- notably Rossby-UNet and CNRM-UNeT --- produce near-zero or slightly negative area-mean signals over New Zealand (-2.56\% and +0.63\% respectively) and Southern Africa (-2.47\% and +3.98\% respectively), indicating a failure to capture even the sign of the domain-mean response in these regions. ResGAN-v2 produces the largest predicted signals among models shown (+6.37\% over the Alps, +23.34\% over Southern Africa, +13.03\% over New Zealand) and in the Southern Africa and New Zealand domains overestimates the reference in certain regions. Across all domains, the RMSE values in the ESD experiment are broadly similar between top- and bottom-ranked models (11--27 \%), underscoring that RMSE alone does not discriminate skill well between models.

In the RCM Emulator experiment (Figure~\ref{fig:rx1day-emul}), the area-averaged climate change signal distribution of the different models is approximately 5 percentage points greater than in the ESD experiment, with more models predicting positive area-mean changes closer in magnitude to the reference. This improvement is evident across nearly all models (e.g., RCMGEM-mv-orog). However, despite this improvement, RMSE values are not markedly lower than in the ESD experiment (10--22\%), and most models continue to underestimate the reference domain-mean change, particularly over New Zealand and Southern Africa where the mean reference signals are largest. There is nonetheless an improvement in the spatial structure of the climate change signal relative to the ESD experiment, with stronger and more positive signals across all regions. Some models predict very weak signals even when trained on the future period; for example, ANN-orog produces near-zero spatial structure in the change field over Southern Africa despite predicting a positive area mean (+4.52\%), reflecting a tendency to spread the signal uniformly rather than concentrate it in regions of enhanced convective organisation. While Figures~\ref{fig:TXx-esd}--\ref{fig:rx1day-emul} focus on the mid-century period, extending the evaluation to the end-of-century (2080--2099) reveals more pronounced underestimation in the ESD experiment, with the RCM Emulator experiment generally improving skill. However, since this period is included in the RCM Emulator experiment's training data, it does not constitute a proper out-of-sample test (Figures~S9--S11).

Overall, these results highlight two findings consistent across all 40 models. First, including future periods during training (RCM Emulator vs.\ ESD) improves the representation of the TXx climate change signal, largely resolving the systematic underestimation seen in the ESD experiment. For Rx1day, training on future data improves the mean magnitude and spatial structure of predicted changes, but does not reduce RMSE in the climate change signal. Second, predicting the climate change signal of precipitation extremes is considerably more challenging than for temperature. The inherently noisy character of the Rx1day signal limits the utility of standard pointwise error metrics as a benchmarking diagnostic --- requiring subjective assessment in addition --- which is why it is not included directly in the scoring framework discussed in Section~\ref{sec:benchmarking}. 


\subsection{Benchmarking}\label{sec:benchmarking}

\begin{figure}
\noindent\includegraphics[width=0.98\textwidth, trim=0 0 0 10mm, clip]{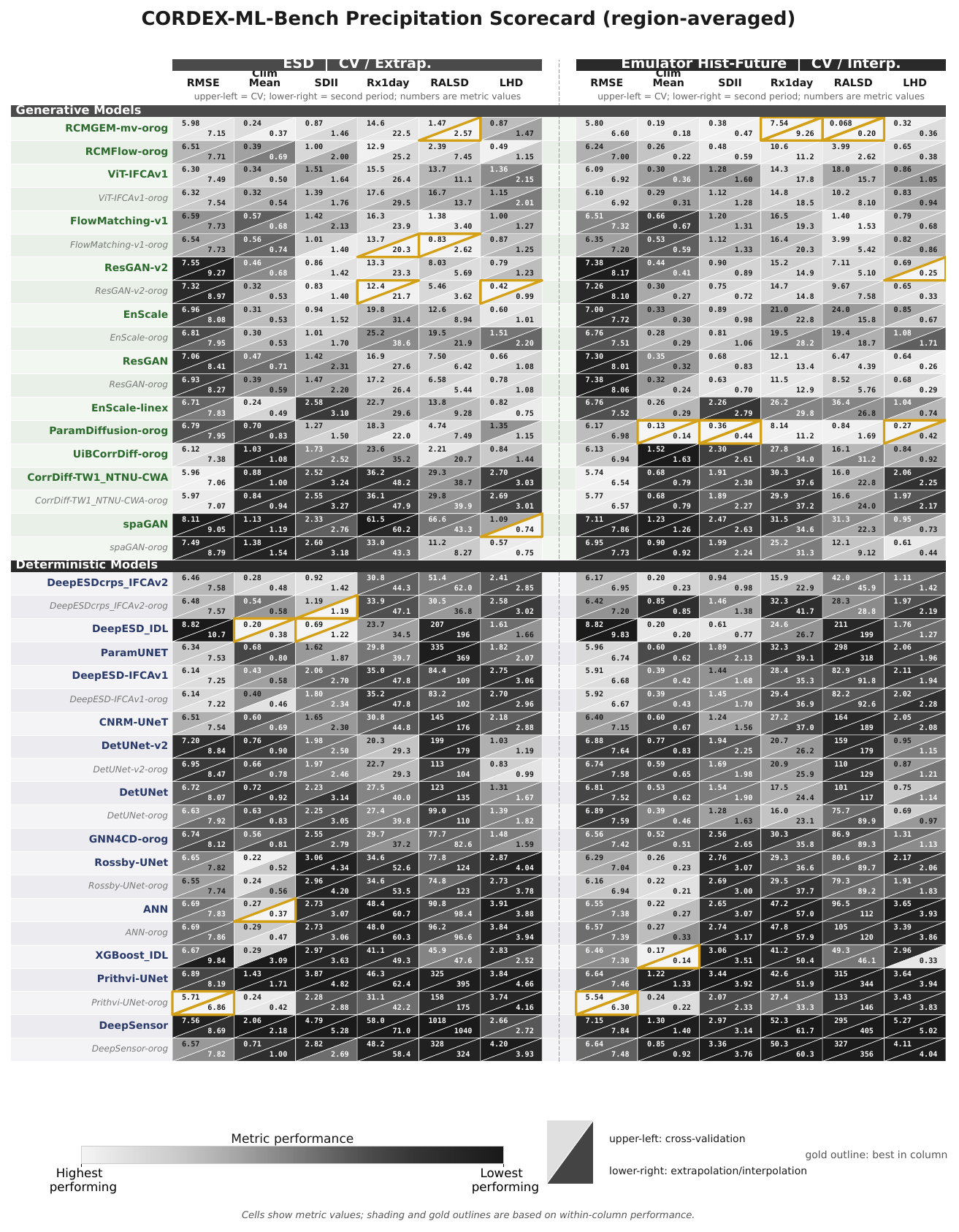}
\caption{Ranking scorecard for precipitation (pr) in the ESD (left) and RCM Emulator experiment (right). Within each scorecard, cell shading indicates relative rank for each metric (column). Each cell is split diagonally, with the shading of the upper-left triangle showing the rank during the historical cross-validation (1981–-2010) period and the lower-right triangle showing the rank during the out-of-sample future period (2041--2060; extrapolation for ESD; interpolation for the RCM Emulator experiment). Lighter shading indicates higher rank (better performance); darker shading indicates lower rank. Values within each triangle are the absolute metric scores. A yellow border marks the best-performing model for each column. Metrics are averaged across three domains (Alps, New Zealand and Southern Africa). Rows are sorted by each model's average rank across all metrics, periods and experiments (ESD and RCM Emulator). The values within each cell are the absolute metric scores.}

\label{fig:score-pr}
\end{figure}

\begin{figure}
\noindent\includegraphics[width=0.98\textwidth, trim=0 0 0 10mm, clip]{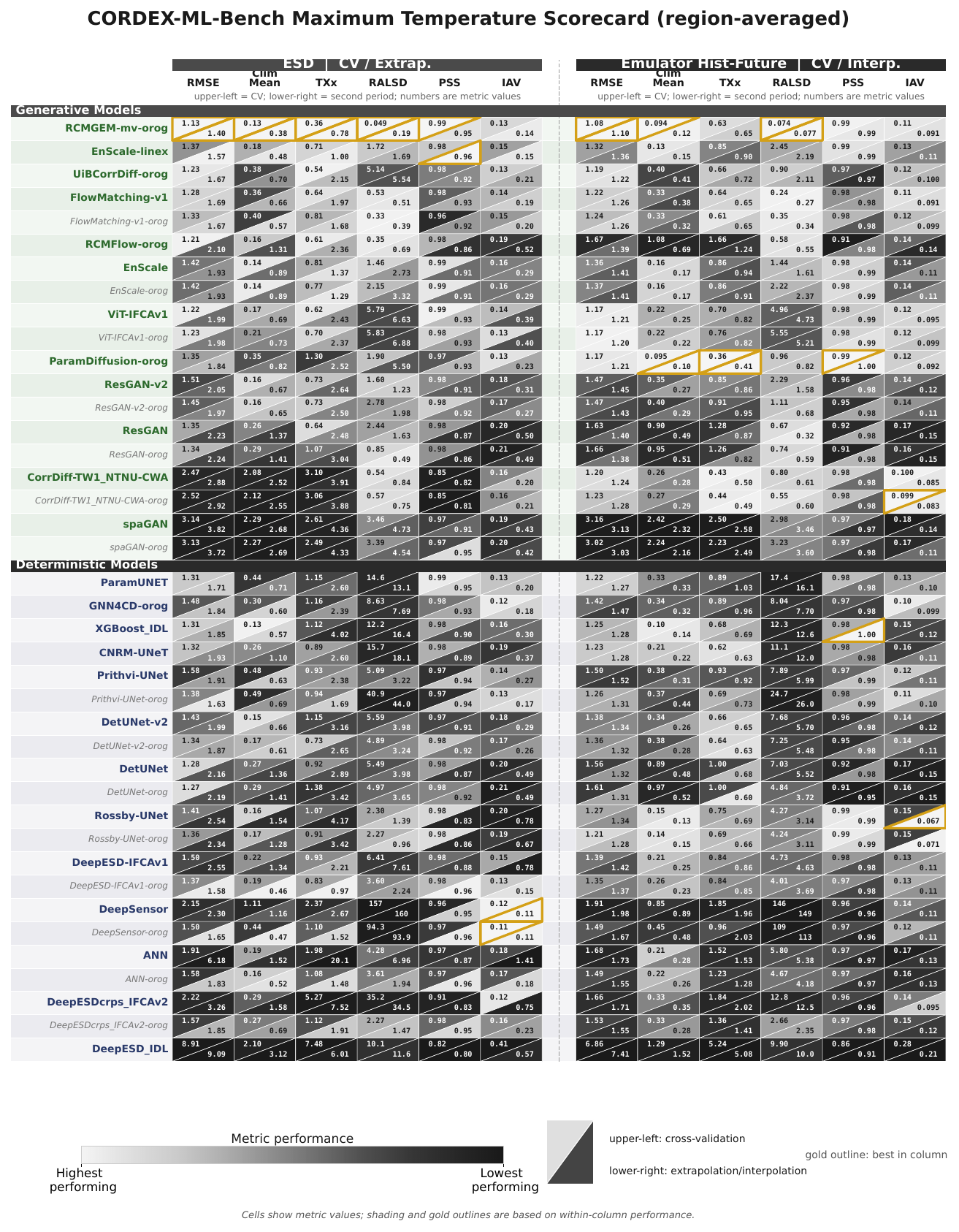}
\caption{As Figure~\ref{fig:score-pr}, but for maximum temperature (tasmax). Metrics differ from precipitation and include TXx RMSE (in place of Rx1day) and interannual variability bias.}
\label{fig:score-tasmax}
\end{figure}

\begin{figure}
\centering
\noindent\includegraphics[width=0.725\textwidth, trim=0 0 0 10mm, clip]{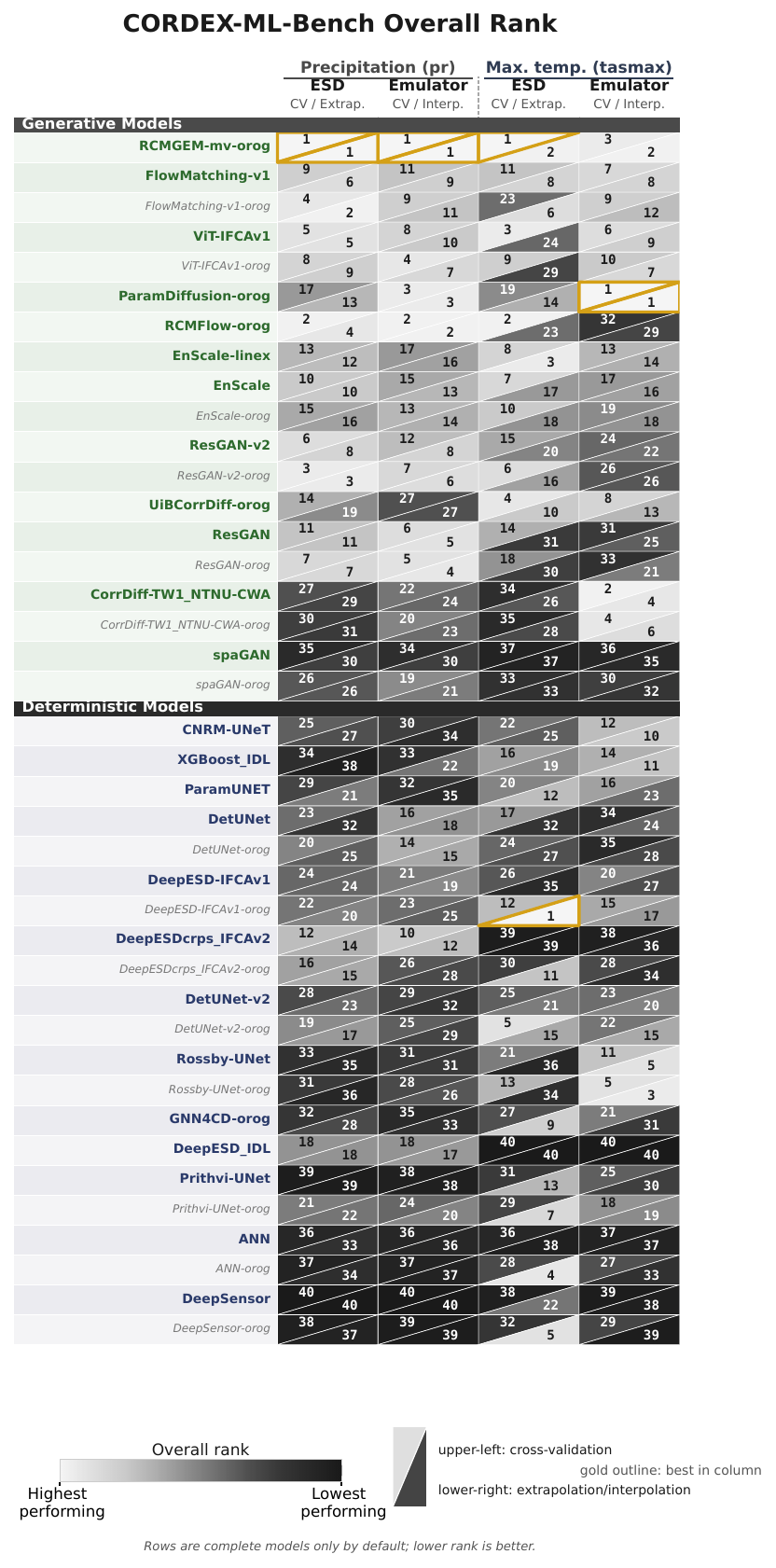}
\caption{Overall rank summary across all experiments and evaluation periods for both precipitation and temperature. Each panel shows the overall rank (aggregated across all metrics and domains) for each model in each of the four experiment--variable combinations: precipitation ESD, precipitation Emulator, temperature ESD, and temperature Emulator. Within each cell, the upper-left triangle shows the cross-validation rank and the lower-right triangle shows the extrapolation or interpolation rank. Models are separated into generative and deterministic groups. Lighter shading indicates higher rank.}
\label{fig:rank-summary}
\end{figure}

Model performance across all metrics, variables, and experiments is summarised in the scorecards shown in Figures~\ref{fig:score-pr}--\ref{fig:score-tasmax}, with an overall rank summary presented in Figure~\ref{fig:rank-summary}. For both variables we report the RMSE, climatological mean RMSE (Clim Mean) and RALSD. To evaluate extremes, we use the Rx1day for precipitation and TXx for maximum temperature. To assess the ability of the models to reproduce the distribution of the target variable, we rely on the PSS for maximum temperature and the LHD for precipitation. This distinction reflects the difference in the distributional nature of these two variables (see Section \ref{subsection:evaluation-metrics} for details). In addition, we specifically address the SDII bias for precipitation and the interannual variability for maximum temperature. The values reported in these scorecards are averaged across all domains; further evaluation within each domain is provided in Supplementary Figures S12–S13. Note, a low rank does not imply that a model has failed in an absolute sense, but rather reflects its standing relative to the full model ensemble evaluated here.

For precipitation in the ESD experiment (Figure~\ref{fig:score-pr}; left), generative models consistently outperform deterministic approaches, particularly on Rx1day, RALSD, and LHD, which assess extremes, spatial variability, and distributional fidelity, respectively. Daily RMSE, widely used in ML, proves a poor proxy for downscaling skill: CorrDiff-TW1\_NTNU-CWA and Prithvi-UNet achieve low RMSE yet score poorly on climatological metrics, while the ResGAN and EnScale variants show the converse. The top-performing models (RCMGEM-mv-orog, RCMFlow-orog, ResGAN-v2, ViT-IFCAv1, and FlowMatching-v1) score consistently well across both cross-validation and future extrapolation periods (Figure~\ref{fig:score-pr}, left panel, rightmost column). Notably, the relative ranking of models is largely consistent across different periods (as indicated by the similarity in shading between triangles within a given cell), suggesting that model performance is stable across periods. In the RCM Emulator experiment (Figure~\ref{fig:score-pr}; right), similar patterns emerge with generally improved scores, as models benefit from training on a future period; here ParamDiffusion-orog additionally proves highly skilful. 

For temperature in the ESD experiment (Figure~\ref{fig:score-tasmax}; left), the performance gap between generative and deterministic models narrows considerably. This may reflect the fact that temperature is spatially smoother and more predictable from large-scale fields \citep[e.g.,][]{doury2023regional}. As a result, the stochastic variability that generative models produce, which is particularly beneficial for precipitation — matters less here, narrowing their advantage over deterministic approaches. RCMGEM-mv-orog remains the top performer across most metrics, while some deterministic models also perform well for certain metrics and periods. A key finding is the large and often inconsistent shift in both relative rank and absolute score between the historical cross-validation and future extrapolation periods (visible as colour changes along the scorecard diagonal), in contrast to the precipitation results. For metrics such as TXx and PSS, individual model scores change substantially between periods, with some models jumping from low to high rank and others dropping in the opposite direction. These apparently random rank reversals indicate that strong cross-validation performance does not reliably predict future-period skill when models must generalise beyond their training distribution. 

In the RCM Emulator experiment (Figure~\ref{fig:score-tasmax}; right), the large shifts in relative rank seen in the ESD setting largely disappear: models that perform well in one period do so consistently in the other, and metrics such as TXx and PSS remain stable across evaluation windows. The top-ranked models are RCMGEM-mv-orog, ParamDiffusion-orog, CorrDiff-TW1\_NTNU-CWA, and ViT-IFCAv1. Deterministic models also achieve stronger scores relative to generative approaches than in the precipitation case. This improved consistency likely reflects the fact that the models are trained on data spanning both historical and future periods, and therefore need not generalise beyond their training distribution. The contrast with the ESD results suggests that the skill degradation observed there stems from extrapolation difficulty, and that access to future-period training data is important for stable performance under future conditions across multiple metrics. 

The overall rank summary (Figure~\ref{fig:rank-summary}) shows model ranks for each variable and evaluation period, consolidated across metrics: for each model, a per-metric rank is computed, averaged across metrics with equal weighting, and then re-ranked to give the final score. RCMGEM-mv-orog is the top-ranked model overall, consistently ranking 1 or 2 across all experiments, variables, and evaluation periods --- a level of consistency not matched by any other model. ParamDiffusion-orog also performs exceptionally well in the RCM Emulator experiment, closely matching RCMGEM-mv-orog. Some models perform well specifically for precipitation --- notably RCMFlow-orog and the ResGAN variants --- but rank in the middle or lower part of the ensemble for temperature. We note that these rankings reflect not only architectural choices but also differences in model selection and hyperparameter tuning, which vary across the different approaches; further tuning and checkpoint selection could significantly influence model skill. Across all models and experiments, domain difficulty follows a consistent ordering: biases are smallest over the Alps, larger over New Zealand, and largest over Southern Africa (Figures~S12–S13 report skill scores per domain). The latter is likely particularly challenging due to its convective-dominated summer environment, which is less strongly coupled to large-scale atmospheric fields. However, this may also partly reflect differences in RCM configuration (e.g., spectral nudging scales) rather than the downscaling task itself \citep{ratnam2013dynamical, soares2024future}, warranting further investigation.

Finally, even the top-ranked models can underestimate the climate change signal in precipitation extremes (Section~\ref{sec:cc_signal}), a limitation not captured by the summary rankings. Model selection for a specific application should therefore consider not only overall benchmark rank but also how well a model represents the climate change signal in the variable and metric of interest. More comprehensive evaluations beyond those presented here will be addressed in future work.

\section{Discussion}

\subsection{The performance--compute trade-off}

Many models in this benchmark were originally developed for purposes other than climate downscaling, and were not specifically optimised for preserving climate change signals or performing well across all evaluation metrics. For example, CorrDiff \citep{mardani2025residual} was designed for fine-scale weather downscaling and excels on metrics such as RMSE and RALSD, while ResGAN and its variants were explicitly optimised for Rx1day and RALSD rather than RMSE \citep{rampal2025reliable}. Also, the ParamDiffusion-orog model was explicitly designed as an RCM emulator rather than a statistical downscaling tool \citep{legasa_regional_2026}, which likely contributes to its comparatively lower performance in ESD experiments. Rankings should therefore be interpreted with this context in mind: targeted development could readily improve skill for many architectures, and the community is encouraged to continue refining and benchmarking these models. Skill alone, however, is only one dimension of practical utility, and training and inference costs also matter. Figure~\ref{fig:skill-compute} illustrates the skill–compute relationship for the ML models in this benchmark, plotting Rx1day and TXx RMSE against inference time. While certain approaches were deliberately optimised (e.g., RCMFlow-orog), most models did not optimise training or inference, and much of this overhead could be reduced with further development. Even so, the figure reveals that inference cost varies by more than 4-5 orders of magnitude across the different approaches. At the low end, inference takes between under one and ten seconds per simulated year for EnScale, ResGAN, and most deterministic approaches. At the high end, RCMGEM-mv-orog requires between 20-40 minutes per simulated year on an A100 GPU \citep[e.g.,][]{addison2026machine}. 

While all approaches remain far more efficient than running RCMs, this four to five order-of-magnitude spread in inference cost has real practical implications in contexts where access to GPUs is limited, particularly for downscaling large initial-condition ensembles of climate projections \citep{aalbers2018local, maher_2021_large}. Computationally efficient approaches such as ResGAN have already been used to downscale ensembles exceeding 15,000 years of projections in under a day on a single A100 GPU \citep{rampal2025downscaling}. Models that are orders of magnitude slower could require weeks or months to complete the same task on a single GPU; with multiple GPUs available, however, parallelising the work would substantially reduce wall-clock time, making even the most expensive models tractable.

In this context, mid- and low-cost generative models offer a compelling balance of skill and computational efficiency. RCMFlow-orog, FlowMatching-v1, ViT-IFCAv1, EnScale, and ResGAN are less skilful than RCMGEM-mv-orog overall, but their inference cost is one to two orders of magnitude lower — making them more tractable for large-scale ensemble applications where computational budget is a practical constraint. A key reason RCMFlow-orog and FlowMatching-v1 are more computationally efficient than RCMGEM-mv-orog is that they require far fewer neural function (flow or diffusion time) evaluations during inference. For example, RCMFlow-orog uses flow matching with an Adams–Bashforth third-order solver, requiring only 25 function evaluations per sample, compared to the hundreds or thousands required by typical score-based SDE samplers such as RCMGEM-mv-orog and ParamDiffusion-orog \citep[e.g.,][]{song2020score}. Non-diffusion generative models such as EnScale and ViT-IFCAv1 go a step further, generating ensemble members in a single forward pass and thus incurring no multi-step cost whatsoever. Adopting similar sampling strategies represents a clear path to reducing inference cost for these slower diffusion-based models, and further cost optimisations may be possible. Moreover, choosing the optimal model involves a trade-off: skill versus computational cost. We therefore recommend that future benchmarking efforts, and ML downscaling studies more generally, continue to report compute cost alongside skill metrics. Note, the results in Figure~\ref{fig:skill-compute} should be interpreted with care. Computational costs are approximate, as the algorithms used different codebases and hardware configurations, making consistent normalisation difficult. 

\subsection{Extrapolation and transferability}\label{sec:discexrtapoation}

\begin{figure}
\noindent\includegraphics[width=1.1\textwidth]{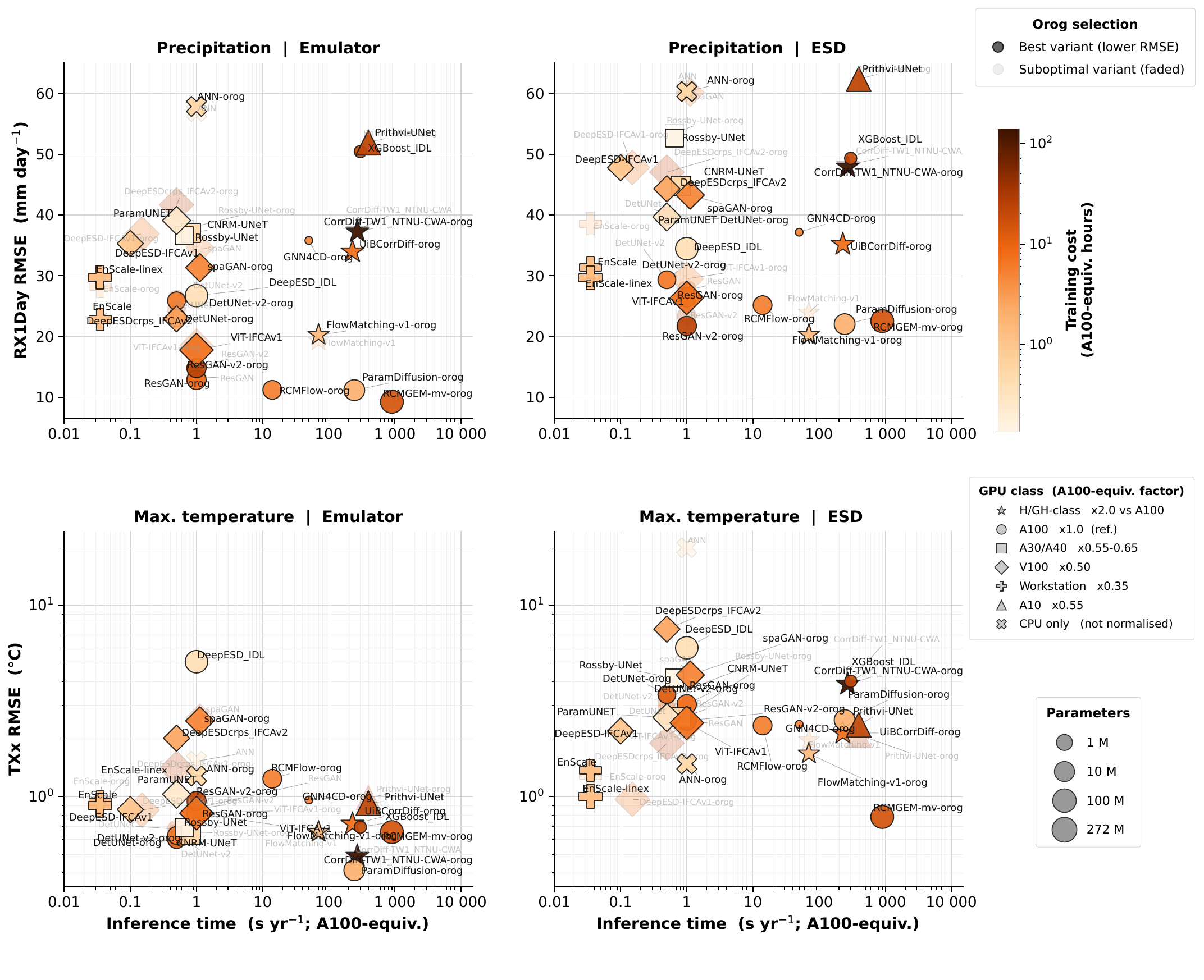}
\caption{Skill--compute relationship across CORDEX-ML-Bench models. The horizontal axis shows inference cost as wall-clock time per simulated year per ensemble member, expressed as a single A100-GPU equivalent. To enable hardware-agnostic comparison, recorded times are scaled by the approximate throughput of each GPU class relative to an A100 (H100/H200/GH200: $\times$2.0; A40: $\times$0.65; A30/A10: $\times$0.55; V100: $\times$0.50; RTX~6000: $\times$0.35; CPU: unchanged). For multi-GPU algorithms, the per-GPU wall-clock time is multiplied by the number of GPUs used; for multivariate models (trained jointly on precipitation and temperature), the cost is halved to give a per-variable equivalent. The vertical axis shows spatially averaged RMSE of annual maximum daily precipitation (Rx1day; top) and annual maximum daily temperature (TXx; bottom) across all three CORDEX domains. Left and right columns show results for the Emulator  and ESD experiments, respectively, evaluated over 2041–2060. Marker colour indicates A100-equivalent training cost; marker size scales with total parameter count; marker shape denotes GPU class (see legend). Where both orographic and non-orographic variants of a model were submitted, the better-performing variant (lower composite RMSE) is shown at full opacity and the weaker variant is faded. }
\label{fig:skill-compute}
\end{figure}

A central aim of CORDEX-ML-Bench is to evaluate extrapolation --- the ability of ML models to generate physically credible fields under climate conditions not present in their training data. The results presented in this study are from the perfect-framework setting, where predictor fields are obtained by coarsening the same RCM used to generate the targets. This isolates the downscaling function from discrepancies between the GCM and RCM atmospheric states (i.e., the imperfect framework). The main finding is that ESD-trained models systematically underestimate mid-century TXx and Rx1day climate change signals, whereas models trained in the RCM Emulator experiment reproduce these signals more accurately. The benefit of including future periods in training is evident in Figure \ref{fig:skill-compute}: models trained on the RCM Emulator configuration have substantially lower errors on average than those trained on historical climate alone, consistent with earlier work \citep{doury2023regional, rampal2024extrapolation, bano2024transferability}.  These results in the perfect framework will thus likely represent lower bounds on the extrapolation errors expected in operational use. When the models trained here are applied in an  imperfect setting (i.e. operationally), where models trained on coarsened RCM fields are driven by raw GCM predictors, skill is likely to degrade further \citep{rampal2025reliable,bano2024transferability,doury2023regional,boe2023simple}. GCM--RCM pairs exhibit systematic differences in the phase and amplitude of large-scale atmospheric variability because RCMs are constrained by GCMs only at lateral boundaries or are nudged to a specific scale (spectral nudging), while their interior states evolve quasi-independently. These inconsistencies are not uniform across variables or regions, and the extent to which they degrade downscaled fields appears to be architecture-dependent \citep{bano2024transferability, kendon2025potential, gonzalez2025deep}. 

A further consideration is resolution: coarsening RCM fields in the perfect framework averages high-resolution variability and leaves an imprint of the fine-scale field on the predictors, so even small differences in spatial variability between coarsened RCM and native GCM predictors may affect model performance. A thorough characterisation of imperfect evaluation, including the interaction between imperfect predictors and out-of-distribution climate states --- is therefore an important subject we plan to address in future work. In particular, a systematic assessment of the climate change signal under imperfect extrapolation, stratified by architecture and region, is needed to determine whether the rankings established under the perfect framework are preserved when models are applied to real CMIP predictor outputs. Previous work suggests that generative architectures may be more robust to predictor perturbations from imperfect conditions than deterministic approaches \citep{rampal2025downscaling, addison2026machine}, but this has not yet been tested at the scale of CORDEX-ML-Bench (i.e., across multiple models and regions). Strategies that make perfect-framework training more representative of inference conditions may prove a useful research direction, and studies such as \citet{aich2026conditional} offer some guidance through spectral smoothing approaches.

A related question is the transferability of ML models to driving GCMs not seen during training, as the present study focuses on a single training GCM. The evaluation matrix in Table~\ref{tab:cordex-ml-bench} includes perfect and imperfect transfer to an independent driving GCM (MPI-ESM-LR for the Alps, NorESM2-MM for South Africa, EC-Earth3 for New Zealand), but for conciseness the analyses presented here are restricted to a single GCM. Cross-GCM evaluation tests a stricter form of generalisation: whether a model has learned a physically meaningful large-scale-to-local mapping, or has instead fit features specific to the training GCM. Preliminary work \citep{rampal2025reliable,addison2026machine,doury2024suitability} has shown that models which appear competitive under the training GCM can degrade substantially when applied to another GCM, so testing this systematically across all 40 models is a natural next step.

\subsection{Architectural and training choices}\label{sec:orography}

Most ML algorithms in the benchmark were provided in paired configurations that differ only in whether a static high-resolution orography field was included as a predictor, yielding an approximately controlled experiment for the importance of explicit topographic information. A systematic analysis of orography versus no-orography pairs is beyond the scope of this paper and will be addressed in future work. For some architectures, incorporating orography consistently improves skill across most metrics (e.g., Prithvi-UNet, ResGAN, Det-UNet, and DeepSensor), as illustrated in Figure \ref{fig:skill-compute}, but this result is not universal. The inconsistency across architectures suggests that orography does not simply act as additional free information; rather, it interacts with each architecture's inductive biases in ways that are not straightforward to predict. Several other architectural and training choices are currently confounded with one another in our analysis. These include the loss function (MSE, asymmetric, quantile, Bernoulli-Gamma negative log-likelihood, or CRPS-spectral), the normalisation strategy for precipitation (log-transform, square-root, percentile scaling, or per-grid-point standardisation), training checkpoints and whether models are trained jointly on precipitation and temperature or independently on each variable. Some of these choices have clear motivations — Bernoulli-Gamma losses explicitly address the zero-inflated character of daily precipitation \citep{cannon2008probabilistic, bano2020configuration}. A controlled study in which a single backbone architecture is trained under different loss functions, normalisation strategies, and variable coupling approaches would allow these choices to be isolated from other sources of inter-model variance.

\subsection{Outlook and Future Extensions of the Benchmark}

The metrics in this first release of CORDEX-ML-Bench are a deliberately chosen core set that covers the main aspects of downscaling evaluation while remaining simple to compute, to encourage rapid uptake in the ML downscaling community. This set is intended to evolve iteratively as new results and user needs emerge from ongoing benchmark activity, with future releases expanding and refining the metrics accordingly. One notable omission from the scorecard reported in Section \ref{sec:benchmarking} is the climate change signal. We excluded it because signal-error metrics are ineffective at discriminating between models when the underlying signal is spatially noisy, as for Rx1day --- although the same metrics were informative for TXx (Section \ref{sec:cc_signal}) --- and more careful metric design is needed before such measures can be meaningfully incorporated.

Beyond the improvements already noted, several research directions are worth pursuing. First, combining a spatially aggregated area-mean change with a pattern-correlation metric would allow evaluation to separately assess biases in the magnitude and spatial distribution of the response --- two distinct error types that are both relevant to capturing the climate change signal. Second, supplementary metrics such as return-period curves and block-maxima or peak-over-threshold distributions would extend evaluation beyond the 20-year mean of Rx1day to the upper tail of the precipitation distribution, which is most directly relevant to impact assessment \citep[e.g.,][]{trenberth2003changing}. Third, the benchmark uses different training data lengths for the ESD and RCM Emulator experiments (20 vs.\ 40 years), which could in principle contribute to differences in performance. We expect this accounts for some, but not all, of the difference: historical cross-validation performance is similar across both experiments for most models (Figures \ref{fig:score-pr}--\ref{fig:score-tasmax}); had ESD scores been systematically lower, a training-length effect would have been the natural explanation. This suggests that the skill improvement of models trained in the Emulator experiment stems from exposure to a warmer future climate during training, rather than from training-length differences --- a finding consistent with results over New Zealand \citep{rampal2024extrapolation}.
A further extension concerns ensemble-based evaluation. Many algorithms are generative and produce ensembles of downscaled fields, yet the current benchmark does not assess ensemble spread or calibration --- properties that would offer valuable insight into the reliability, and added value, of probabilistic outputs \citep[e.g.,][]{addison2026machine, rampal2025reliable, schillinger2025enscale}. Proper scoring rules such as the continuous ranked probability score (CRPS), spread--skill diagrams, and rank histograms are standard tools in probabilistic forecasting \citep{gneiting2007strictly} and should be integrated into future benchmark releases.

Several extensions of CORDEX-ML-Bench are planned for subsequent releases, summarised here in approximate order of priority. The most important next step is the extension to imperfect evaluation discussed in Section \ref{sec:discexrtapoation}, for which the required experiments are already available. This is particularly important because it directly reflects the operational use case that motivates this work, for both ESD and RCM emulation strategies. This extension covers evaluation with raw GCM forcing, transfer to independent GCMs unseen during training (e.g., training on CNRM-CM5, testing on MPI-ESM-LR), and end-of-century conditions (2080–2099). Future phases will revisit these questions more thoroughly and extend to others, including transferability across RCM/GCM simulation pairs and the importance of orography as a co-variate across different ML architectures. Second, expansion to additional CORDEX domains --- particularly tropical, semi-arid, and monsoon-dominated regions, as well as high-latitude domains --- would test the geographic generality of the conclusions drawn here. Third, extension to additional variables, including near-surface wind, humidity, radiation, and sub-daily precipitation --- would support a broader range of impact applications, including hydrology, renewable energy, and urban heat assessment \citep{johannsen2024deep}. Sub-daily resolution is a particular priority, as diurnal-cycle errors are a well-documented limitation of climate models \citep{christopoulos2021assessing, dai2004diurnal, flato2014evaluation}, and data-driven downscaling has shown early promise at these time-scales \citep{johannsen2024deep,glawion2023spategan,glawion2025global}. Fourth, convection-permitting (kilometre-scale) targets are the logical endpoint of this progression \citep[e.g.,][]{addison2026machine}: the factor-of-8 resolution ratio currently evaluated (approximately 2$^\circ$ to 10 km) is modest by the standards of operational convection-permitting downscaling, and scaling to the full ratio required by kilometre-scale targets will challenge current architectures in new ways \citep{kendon2025potential, rampal2024enhancing}.

\section{Conclusions}

To our knowledge, CORDEX-ML-Bench is the first coordinated multi-domain, multi-architecture benchmark for data-driven regional climate downscaling. It provides three primary contributions. First, it aligns with the CORDEX to assess model skill under operational conditions relevant to climate-impact assessments, serving as a foundational release intended for community extension. Second, it evaluates over 40 diverse models—ranging from traditional machine learning baselines to advanced generative architectures—across three regions with different characteristics, enabling systematic, architecture-level comparisons. Third, it evaluates model skill both within and beyond training distributions, distinguishing between cross-validation and out-of-sample performance (extrapolation for ESD models and interpolation for RCM emulators).
Beyond these contributions, the benchmark was designed to address two key questions: first, how well do models trained on historical climate extrapolate to future conditions, and does including future training data improve out-of-sample performance; and second, are certain architectures consistently superior across regions and evaluation criteria? The remainder of this section summarises our findings on each.

On the first question, an important result of this benchmark concerns the ability of models to extrapolate beyond their training distribution. When trained exclusively on a historical period of simulation (the ESD experiment), nearly all 40 models systematically underestimate mid-century climate-change signals for temperature (TXx) and precipitation (Rx1day) extremes. This systematic bias across architectures and regions reinforces recent literature highlighting the potential limitations of using purely historical data to extrapolate non-stationary climate relationships \citep{rampal2024enhancing, kendon2025potential}. Furthermore, because our simulation-based experiments are idealized, we anticipate these extrapolation challenges would be amplified in noisier, observation-based, real-world settings which often use coarse-resolution reanalysis as predictors and gridded observations as targets \citep{rampal2024enhancing}. In such settings, the choice of observational data set itself constitutes an additional source of uncertainty, substantially modifying the downscaled climate change signal, particularly for extremes \citep{reyes-elgueta_observational_2026}.

Training on a combined historical and future period largely resolves the underestimation of the TXx climate change signal across all three domains. We attribute this primarily to exposure to future climate conditions during training, with the larger training set (40 vs.\ 20 years) playing a secondary role --- consistent with previous studies \citep{chadwick2011artificial, rampal2024extrapolation}. For Rx1day, the combined training period reduces but does not eliminate underestimation of the climate-change signal, and this bias is consistent across virtually all architectures. This likely reflects the inherent difficulty of learning the climate-change response of precipitation extremes, which is further challenged by the low frequency of extreme events in training data. Additionally, these results should be interpreted in light of the benchmark's experimental design, where the validation approach was not constrained to climate-change diagnostics; future work could explore incorporating such metrics into model selection and hyperparameter tuning.

As for the second question, across both experiments and all three pilot domains, generative model families—specifically score-based diffusion and flow-matching approaches—achieve the highest overall skill. They consistently outperform deterministic architectures in capturing spatial variability (RALSD) and extreme precipitation (Rx1day), though this advantage narrows for maximum temperature, where several deterministic models remain competitive. A key finding is that pixelwise daily RMSE is a poor proxy for overall downscaling skill. Models with low RMSE often rank poorly on climatological and extreme-value metrics, and vice versa —highlighting the importance for a multi-metric evaluation framework. While our analysis provides a comprehensive overview of model skill, important avenues for future work remain. These include process-based validation (e.g., evaluating cyclone-induced rainfall) and, for generative models, assessment of ensemble dispersion and calibration. Furthermore, it is important to evaluate model skill in predicting events rarer than those considered here (e.g., 1-in-100-year extremes). This is particularly important because the computational efficiency of data-driven models enables the generation of large ensembles, allowing us to sample and study these rare climate events at fine spatial scales.

Although all approaches are far more efficient than dynamical downscaling, inference costs span over two orders of magnitude across models. Flow-matching and GAN-based architectures offer favourable skill-to-compute ratios, making them well-suited to regions with limited GPU infrastructure, while the higher cost of more expensive models can be readily mitigated through multi-GPU parallelisation where resources permit. We recommend that future benchmarking efforts report inference cost alongside skill metrics as a standard evaluation axis, following the approach adopted in WeatherBench 2 \citep{rasp2024weatherbench}.
Future extensions of CORDEX-ML-Bench will expand the benchmark by evaluating models in the imperfect setting (applying models trained on coarsened RCM predictors to raw GCM fields), further analysing the ESD experiment to better characterise model behaviour under climate change conditions, incorporating new CORDEX domains (tropical, high-latitude, and monsoon regions), and targeting sub-daily, convection-permitting kilometer-scale resolutions. We view  CORDEX-ML-Bench as an early step toward building trust in these methods. Realising that goal will require sustained effort beyond the benchmark itself, in particular, systematic comparisons against observations and closer engagement with the impacts-modelling and climate-services communities, to ensure that ML downscaling products are evaluated, deployed, and interpreted in ways that genuinely serve end users.

\section*{Open Research}
The CORDEX-ML-Bench benchmark dataset and code used in this study is publicly available on Zenodo
under a Creative Commons licence \citep{Rampal2026_CORDEX_ML_Bench_companion}, provided as regional
NetCDF files in compressed archives (approximately 30~GB total; approximately 5~GB per
domain) at daily temporal resolution.
The benchmark framework, data loading utilities, and training infrastructure are available
at \url{https://github.com/WCRP-CORDEX/ml-benchmark}; individual model repositories are
linked therein, and additional model submissions are actively encouraged.
All evaluation metrics and scorecards were produced using the CORDEX-ML-Bench evaluation
suite, available at \url{https://github.com/jgonzalezab/cordex-bench-eval}. Code for the models trained in the benchmark is listed in Supplementary Table T3. 

\section*{Supplementary Material}
Supplementary Material can be found in https://zenodo.org/records/20985924

\section*{Author Contributions}
N.R., J.G.-A., and J.M.G. planned and designed the benchmark, and coordinated its development with all contributing members. N.R. and J.G.-A. processed the data, with support from J.S. (Steinkopf), C.H., and F.E.\\
N.R., J.G.-A., H.A., V.B., S.D.G., J.O-D., A.D., R.F.-F., L.G., M.I., H.K.L., M.N.L., M.O., J.P., M.S.J.R., M.S. (Schillinger), S.S. (Sharma), W.T., J.-B.T., R.T., K.-C.W., and T.W. were involved in training ML models and ML algorithm design.
M.L.B., B.B., E.C., F.E., P.B.G., C.H., A.O., M.S.J.R., P.M.M.S., S.S. (Sobolowski), J.S. (Steinkopf), J.B-M, Y.-C.W., P.A.G.W., T.W. (Wetherell), and M.W. contributed to supervision of model training, development of evaluation metrics, and reviewing and editing of the manuscript.

\section*{Conflict of Interest}
The authors declare no conflicts of interest.

\acknowledgments

Authors N.R.\ and P.B.G.\ acknowledge support from the New Zealand Ministry of Business, Innovation and Employment (MBIE) Endeavour Fund Smart Ideas programme, grant NIW2504.
J.G.-A., J.B.-M., J.M.G., S.S (Sobolowski), H.A. and J.O.-D acknowledge support from the Copernicus Climate Change
Service (C3S), under contract C3S2\_384, implemented by ECMWF on behalf of the European
Union.
H.A.\ and P.A.G.W.\ were supported by Natural Environment Research Council grant
NE/Z000076/1.
M.L.B.\ acknowledges support from CREATOR-ANR-25-CE56-3663.
V.B., E.C., and A.D.\ acknowledge support from Horizon Europe project Impetus4Change
(I4C; grant 101081555). 
J.O.-D\ is funded by a European Union Horizon 2020 Marie Sk\l{}odowska-Curie Action
(grant 101151904).
R.F.-F.\ and M.I.\ acknowledge support from the Bolin Centre for Climate Research and
the Horizon Europe AI4PEX project (grant 101137682).
L.G.\ acknowledges Helmholtz Research Field Earth and Environment support through the
Innovation Pool Project ACTUATE.
H.K.L.'s work was carried out at the Jet Propulsion Laboratory, California Institute
of Technology, under a contract with the National Aeronautics and Space Administration
(80NM0018D0004).
M.N.L.\ received funding from Agence Nationale de la Recherche~-- France 2030 as part
of the PEPR TRACCS programme under grant numbers ANR-22-EXTR-0005 and ANR-22-EXTR-0011,
with computing and storage resources from GENCI at IDRIS on Jean~Zay.
M.O.\ is funded by the AI4Science PN070500 fellowship within the Generaci\'{o}n~D
initiative, funded by the European Union NextGenerationEU funds through PRTR.
A.O., M.S.J.R., S.S., and M.W.\ acknowledge support from UKRI/NERC; A.O., S.S., and
M.W.\ were funded by the grant Drivers and Impacts of Extreme Weather Events in
Antarctica (ExtAnt; NE/Y503307/1), and A.O.\ and M.S.J.R.\ also acknowledge the NERC
National Capability International grant SURface FluxEs In AnTarctica
(SURFEIT; NE/X009319/1).
J.P.\ was funded by the HClimRep project as part of the Helmholtz Foundation Model
Initiative, with collaboration enabled through a research stay supported by Karlsruhe
House of Young Scientists.
M.S.\ is part of SPEED2ZERO, a joint initiative co-financed by the ETH Board.
P.M.M.S.\ and R.T.\ acknowledge funding from FCT, I.P./MCTES through national funds
(PIDDAC): LA/P/0068/2020, UID/50019/2025, and European Union NextGenerationEU projects
UID/PRR/50019/2025 and UID/PRR2/50019/2025.
J.-B.T., K.-C.W., and Y.-C.W.\ were funded by the National Science and Technology
Council, Taiwan (grants 113-2111-M-003-005, 114-2111-M-003-007, and
112-2923-M-001-003-MY4); J.-B.T.\ and K.-C.W.\ also acknowledge the Central Weather
Administration, Taiwan.
The CCAM simulations contributing to this benchmark for the South Africa domain were produced through TIPPECC
(Climate change information for adapting to regional tipping Points), part of the
Southern African Science Services Centre for Climate Change and Adaptive Land Management
(SASSCAL) 2.0 Research Programme; C.H.\ is additionally supported by the Wits-Nedbank
Chair in Climate Modelling via donation from the Nedbank Eyethu Community Trust.
This work was facilitated by the CORDEX Machine Learning Task Force, established under
the World Climate Research Programme Coordinated Regional Climate Downscaling
Experiment.
The authors thank the modelling groups, data centres, and contributing research
institutions whose simulations and model submissions made this community benchmark
possible.

\bibliography{cordex_ml_bench_refs}

\end{document}